\documentclass[prd,aps,floats,preprintnumbers,preprint]{revtex4}

\usepackage{graphicx, epsfig}
 
\newcommand{\eqr}[1]{Eq.~{(\ref{#1})}}
\newcommand{\be}{\begin{equation}}
\newcommand{\ee}{\end{equation}}
\newcommand{\bea}{\begin{eqnarray}}
\newcommand{\eea}{\end{eqnarray}}

\begin{document}

\preprint{SU-GP-03/4-1, MAD-TH-03-1, CERN-TH/2003-104}

\setlength{\unitlength}{1mm}

\title{Running of the Scalar Spectral Index \\
from Inflationary Models}
 
\author{Daniel J.H. Chung$^{(1,2)}$, Gary Shiu$^{(2)}$ and Mark Trodden$^{(3)}$}
\affiliation{
$^{(1)}$CERN Theory Division, CH-1211 Geneva 23, Switzerland. \\
$^{(2)}$Department of Physics, University of Wisconsin, Madison, WI 53706, USA.\\
$^{(3)}$Department of Physics, Syracuse University, Syracuse, NY 13244-1130,
USA.
}
 
\begin{abstract}
The scalar spectral index $n$ is an important parameter describing
the nature of primordial density perturbations. Recent data, including that from the
WMAP satellite, show some evidence that the index runs
(changes as a function of the scale $k$ at which it is measured) from
$n>1$ (blue) on long scales to $n<1$ (red) on short scales.  We
investigate the extent to which inflationary models can accommodate
such significant running of $n$.  We present several methods for
constructing large classes of potentials which yield a running spectral index.  We
show that within the slow-roll approximation, the fact that $n - 1$
changes sign from blue to red forces the slope of the
potential to reach a minimum at a similar field location.  We also
briefly survey the running of the index in a wider class of inflationary
models, including a subset of those with non-minimal kinetic terms.
\end{abstract}
 
 
\maketitle
 
\section{Introduction.}
\label{introduction}
Primordial perturbations from inflation currently provide our only
complete model for the generation of structure in the universe. It is
commonly stated that a generic prediction of inflationary models is a
scale-invariant spectrum of adiabatic perturbations, characterized by
a scalar spectral index $n$ that obeys $n-1=0$. However, this
statement is only true for very special spacetimes like a pure de
Sitter spacetime, which does not describe our cosmological history.
For nearly all realistic inflationary models, the value of $n$ will
vary with the wave-number $k$.

Typically, since $n-1\simeq 0$ on the scales probed by the cosmic microwave
background (CMB), the deviations from a constant $n$
must be small.  Nevertheless, increasingly accurate cosmological
observations provide information about the scalar spectral index on
scales below those accessible to anisotropy measurements of the
CMB. Over such a wide range of scales, it is entirely possible that
$n$ will exhibit significant {\it running}; a value that depends on
the scale on which it is measured. Such running is quantified by the
derivative $dn/d\ln k$ and, in fact, recently released
data~\cite{Bennett:2003bz} from the WMAP satellite indicates that
\begin{equation}
\label{WMAP}
\frac{dn}{d\ln k} = -0.03^{+0.016}_{-0.018} \ .
\end{equation}
Furthermore, as pointed out by the WMAP analysis of
\cite{Peiris:2003ff}, there is some indication that the spectral index
quantity $n -1$ runs from positive values on long length scales to
negative values on short length scales (positive to negative within
about 5 e-folds).

One must bear in mind that the WMAP analysis of $dn/d\ln k$ may
require modification \cite{Seljak:2003jg} and that the statistical
significance of this result is questionable
\cite{yunwang,Barger:2003ym}.  Furthermore, it has been argued that
the necessary reionization history may be in conflict with large
running \cite{haimanetal}.  Nonetheless, since a running as large as
the central value of Eq.(\ref{WMAP}) is not inconsistent with the
present data, and since it is difficult to produce in common
realizations of inflation, it is worthwhile exploring what such
observational data would imply for inflationary models.

In this paper we investigate the extent to which a scalar spectral
index that runs significantly negatively can be accommodated within
inflationary models, within the first order slow-roll approximation~\footnote{The first model with this desirable feature was 
proposed in \cite{LindeRiotto} although, in general, the running of the
spectral index from blue to red within about 5-efolds is not easy
to achieve.}.
One generic result of our investigation is that within the slow-roll
approximation, there must be an approximate local minimum in the slow
roll parameter $\epsilon$ (which measures the slope of the potential)
if the spectral index is to run from $n>1$ (blue) on long length
scales to $n<1$ (red) on short length scales.  Hence, if
observational evidence continues to show $n-1$ running from
positive to negative, one can infer that the inflaton potential
naturally has a locally flat part in the field values of observational
interest.  Furthermore, the fact that $dn/d\ln k$ is large in this
same field range and the fact that this large value cannot be
sustained for many e-folds (since inflation will then end too quickly)
together indicate that there must be a flat ``bump''-like structure in
the potential.

Unless the bumps have some kind of singular structure, as we will
further explain, they generically lead to $dn/d\ln k<O((n-1)^{2}$. If
$n-1$ never changes sign, then a large $dn/d\ln k$ can be
accommodated without a bump, even though the initial field value must
still be a special point in the potential.  Indeed, the existence of such a special
feature in the potential must be generic since as we will show, a
negative running of the order of Eq.(\ref{WMAP}) requires a large
third (field-)derivative while maintaining small first and second
derivatives~\footnote{A classification of running models was carried out
in~\cite{Hannestad}, in which the need for a large third derivative was
also noted.}.

Our second major set of results concerns methods to construct
potentials with large running and/or potentials in which the scalar
index runs from blue on long length scales to red on short length
scales.  This task is generically nontrivial due to the difficulty of achieving 
around 60 e-foldings of inflation after attaining a large
$dn/d\ln k$.  We develop two formalisms to generate such viable
inflationary potentials.  Through one of our formalisms we show that
having a blue spectrum on long length scales generically helps to
achieve a large number of e-foldings after having attained a large
$dn/d\ln k$.  In this sense, having a blue spectrum with large
$dn/d\ln k$ is not as ``singular'' as having a red spectrum with large
$dn/d\ln k$.

The first method of construction, which we will call the singular
method for reasons that will become apparent (although there need not
be any true singularities), is suitable for generating arbitrarily
large $dn/d\ln k$ without upsetting slow-roll.  
However, this method
of construction does not give rise to a change in the sign of $n -1$ 
within about 5 efolds nor does it yield a blue spectrum on large scales.  The
second method of construction, which we will call the index method,
gives the potential as a functional of any reasonable ansatz for $n
-1$.  This index method is generic, covering virtually all slow
roll models that have a blue spectrum on large length scales.  Within
this formalism, we can easily construct models in which $n-1$ goes
from blue to red.  This method may also be combined with other
formalisms.  

In the appendix, we survey a variety of inflationary models in the
literature and show how strongly the spectral index can run in each
case.  The survey will allow the reader to gain intuition about the
difficulty in achieving strong running of the spectral index as well
as the difficulty in having a spectrum whose index runs from blue to
red.  Although there have been several recent works concerned with the
construction of potentials that have the strong running properties
that we are concerned with
\cite{Feng:2003mk,Lidsey:2003cq,Kawasaki:2003zv,Huang:2003zp}, for
practical reasons we have reviewed only \cite{Feng:2003mk}.
We apologize to the other authors whose work we do not review in the
appendix.  

This paper is organized as follows. In the next section, we establish
our notation and list the well-known constraints on inflationary
models.  In particular, we show that a large third derivative of the
inflaton potential is required to generate a large running of the
spectral index. In section~\ref{sec:canonrecipe}, we develop our first
simple formalism, the singular method, to generate potentials with
large running of the spectral index.  In the section that follows
(section~\ref{sec:canonexamples}), we employ our formalism to obtain
some models of inflation with large running. 
Complementing these
sections on the singular method is section~\ref{sec:normalization},
where we show how nonminimal kinetic terms can aid in making the
construction look more natural.  
In section~\ref{sec:functional}, we
introduce the index method by
deriving an expression for the potential as a functional of any
reasonable spectral index function and in section ~\ref{sec:bluetored}
we use this functional to generate
potentials which give running from blue to red.  
In section~\ref{sec:newconsistency}, we discuss the
result that $\epsilon$ generically has an approximate local
minimum near the point at which $n-1$ changes sign.  Finally we summarize and
offer our conclusions.

\section{Notations and constraints}
\label{sec:notation}
Following Garriga and Mukhanov \cite{GarrigaMukhanov},
let us consider the most general local action for a scalar field
coupled to Einstein gravity, which involves at most first derivatives
of the field,
\begin{equation}
S = - \frac{1}{16 \pi G} \int \sqrt{-g} R d^4 x + \int \sqrt{-g} ~p(X,\phi)
dx^4~,
\label{action}
\end{equation}
where $R$ is the Ricci scalar
and
\begin{equation}
X = \frac{1}{2} g^{\mu \nu} \partial_{\mu} \phi \partial_{\nu} \phi \ .
\end{equation}
The Lagrangian density for the scalar field is denoted by $p$ because
it plays the role of pressure \cite{k-inflation} in cosmological applications. This action reduces
to the usual inflationary model with canonical kinetic term when
$p(X,\phi)=X - V(\phi)$, but it
also describes the more general case in which the kinetic term
is non-canonical. For completeness, we will derive the properties 
of the power spectrum for the general form of $p(X,\phi)$ and apply our results 
to some special cases.

For an action of the form in Eq.~(\ref{action}), the energy density is given
by
\begin{equation}
{\cal E} = 2 X p_{,X} - p \ ,
\end{equation}
where $p_{,X}$ denotes partial derivative of $p$ with respect to $X$.
The ``sound speed'' $c_s$ is then
\begin{equation}
c_s^2 \equiv \frac{p_{,X}}{{\cal E}_{,X}} = \frac{{\cal E}+p}{2X {\cal E}_{,X}} \ ,
\end{equation}
and the power spectrum for the scalar fluctuations is \cite{GarrigaMukhanov}
\begin{equation}
P_k^{\zeta} = \frac{16}{9} \frac{G_N^2 {\cal E} }{c_s
\left(1+p/{\cal E} \right) } \ .
\end{equation}

The spectral index for the scalar mode is given by~\cite{GarrigaMukhanov}
\begin{equation}
n - 1 \equiv \frac{d \ln P_k^{\zeta}}{d \ln k} =
-3 \left(1 + \frac{p}{\cal E} \right) - \frac{1}{H} \frac{d}{dt} \left[
\ln \left( 1 + \frac{p}{\cal E} \right) \right]
 - \frac{1}{H} \frac{d}{dt} \left( \ln c_s \right) + \dots \ ,
\end{equation}
where the quantities on the right-handed side are evaluated at horizon
crossing. Here, we keep only terms that are lowest order in the 
slow-roll parameters $(1+p/{\cal E})$ and  $H^{-1} d \ln c_s/dt$.

This allows us to derive an expression for the main quantity of interest to us, the running spectral index
\begin{eqnarray}
\frac{dn}{d\ln k} = &-& \frac{3}{H} \frac{d}{dt} \left(1 + \frac{p}{\cal E} \right) - \frac{1}{H^2} \frac{d^2}{dt^2} \left[ \ln\left(1 +  \frac{p}{\cal E} \right)
\right] - \frac{1}{H^2} \frac{d^2}{dt^2} \left( \ln c_s \right) \nonumber \\
&-&\frac{1}{H}\left\{\frac{d}{dt}\left[ \ln \left(1 +  \frac{p}{\cal E} \right)\right]\right\}\frac{d}{dt}\left(\frac{1}{H}\right)-\frac{1}{H}\left[\frac{d}{dt}\left( \ln c_s \right)\right]\frac{d}{dt}\left(\frac{1}{H}\right) \ .
\end{eqnarray}

Finally, it is also useful to note that the ratio of the tensor fluctuations $P^{h}$
to the scalar fluctuations $P^{\zeta}$ \cite{GarrigaMukhanov} is given by
\begin{equation}
r \equiv \frac{P^h}{P^{\zeta}} = 24 c_s (1+\frac{p}{{\cal E}})= - 8 c_s n_T \ ,
\end{equation}
where $n_T$ is the tensor spectral index. In principle, the consistency
condition $r=-8n_T$ can be violated for models with non-canonical terms, since
$c_s$ can differ from unity \cite{GarrigaMukhanov,ShiuTye}.

\subsection{The Canonical Limit}
Let us first apply the above general formulae to the standard case of
slow-roll inflation.  In this case,
\begin{equation}
p(X,\phi) = X - V(\phi) \ .
\end{equation}
Therefore, $c_s=1$ and
\begin{equation}
1 + \frac{p}{\cal E} \sim \frac{2X}{V} = \frac{\dot{\phi}^2}{V}
= \frac{V^{\prime 2}}{9H^2V} = \frac{1}{3} M_P^2 \left( \frac{V^{\prime}}{V}
\right)^2 \ .
\end{equation}
Furthermore,
\begin{equation}
\frac{d}{dt} \ln (1 + \frac{p}{\cal E}) =
\frac{2 V}{V^{\prime}} \left[ \frac{V^{\prime \prime}}{V} - \left(\frac{V^{\prime}}{V} \right)^2 \right] \dot{\phi}
= - \frac{2}{3} \frac{V}{H} \left[ \frac{V^{\prime \prime}}{V} - \left(
\frac{V^{\prime}}{V} \right)^2 \right] \ .
\end{equation}

In analyzing the inflationary dynamics driven by $\phi$ it is
convenient to define the conventional slow-roll parameters
\cite{Liddle:1992wi,Liddle:1994dx,Lidsey:1995np}
\begin{equation}
\epsilon \equiv \frac{1}{2}M_{p}^{2}\left( \frac{V'}{V}\right) ^{2} \ ,
\end{equation}
\begin{equation}
\eta \equiv M_{p}^{2}\left( \frac{V''}{V}\right) \ ,
\end{equation}
where the primes denote differentiation with respect to $\phi$ and
$M_p \equiv 1/\sqrt{8 \pi G}$.  Inflation occurs if $\epsilon \ll 1$
and $|\eta|\ll 1$.  In addition, we define a third  parameter
related to the third derivative of the potential
\begin{equation}
\xi \equiv M_{p}^{4}\left(\frac{V'V'''}{V^{2}}\right) \ ,
\label{eq:xidefine}
\end{equation}
which is important for the running of the spectral index.

In terms of the slow-roll parameterization, the power spectrum
becomes
\begin{equation}
P_k^{\zeta} = \frac{16}{3\pi^2 \epsilon}\left(\frac{V}{M_p^4}\right) \ ,
\end{equation}
to leading order.  We also recover the well-known expressions for the
spectral index
\begin{equation}
n - 1 = 2 \eta - 6 \epsilon,
\label{eq:nsmin1slowcanon}
\end{equation}
its derivative 
\begin{equation}
\frac{dn}{d \ln k} = -2 \xi + 16 \epsilon \eta - 24 \epsilon^2 ,
\label{eq:runningslowcanon}
\end{equation}
and \footnote{There is a sign error for the $dn/d\ln k$ expression in
Ref.~\cite{Lyth:1998xn}.}
the tensor to scalar ratio
\begin{equation}
r = 16 \epsilon.
\end{equation}

\subsection{Non-Canonical Kinetic Terms}
Let us now consider a second interesting special case, in which $p(X,\phi)$ takes
the following form
\begin{equation}
\label{generalL}
p (X,\phi) = Z(\phi) X - V(\phi) \ ,
\end{equation}
where the potential $V(\phi)$ and the function $Z(\phi)$ are general
functions.  This form of $p(X,\phi)$ can arise, for example, from quantum
corrections to the kinetic term, which yield $Z(\phi)=1 + c g^2 \ln \phi$ (where $c$ is a
constant, and $g$ is a coupling constant). A similar
action arises in brane inflationary models
\cite{ShiuTye,DvaliTye} due to a velocity-dependent potential between
D-branes.  Finally, note that this form of the nonminimal kinetic term
can always be brought back to the canonical form (at least over a
finite field region) by an appropriate field redefinition, as we will discuss in detail later.
 
The energy density is
\begin{equation}
{\cal E} =  Z(\phi) X + V(\phi) \ ,
\end{equation}
and, since $c_s^2=1$ in this case, the slow-roll parameters depend only on $(1 + p/{\cal E})$, given by
\begin{equation}
1 + \frac{p}{\cal E} = \frac{2 Z X}{ZX+V} \ .
\end{equation} 

To determine the classical background of $X$, let us
consider the equation of motion for $\phi$ 
\begin{equation}
Z \left( \ddot{\phi} + 3 H \dot{\phi} \right) - \frac{1}{2} Z^{\prime} \dot{\phi}^2  + V^{\prime} = 0 \ .
\end{equation}
For slow-roll inflation,  $\ddot{\phi} << 3 H \dot{\phi}$ and
the potential energy dominates. It is thus reasonable to assume
that $Z^{\prime} \dot{\phi}^2 /2<< V^{\prime}$ and
therefore~\cite{ShiuTye}
\begin{equation}
\dot{\phi} = -\frac{V^{\prime}}{3HZ} \ .
\end{equation}
Hence, the classical background value of $X$ is
\begin{equation}
X = \frac{1}{2} \dot{\phi}^2 =\frac{1}{3} \frac{\epsilon V}{Z^2} \ ,
\end{equation}
where $\epsilon$ is the slow-roll parameter defined previously.
Therefore, we may write
\begin{equation}
1 + \frac{p}{\cal E} = \frac{2}{3} \frac{\epsilon}{Z} \left( 1 - \frac{1}{3} \frac{\epsilon}{Z} 
+ \dots \right) \ ,
\end{equation}
and
\begin{equation}
\frac{d}{dt} \ln \left( 1 + \frac{p}{\cal E} \right)
= \frac{1}{\epsilon} \frac{d \epsilon}{d \phi} \dot{\phi}
- \frac{1}{Z} \frac{dZ}{d \phi} \dot{\phi} + \dots \ ,
\end{equation}
where we have used
\begin{equation}
\frac{d \epsilon}{d \phi} = \left( \frac{V^{\prime}}{V} \right) \left( \eta - 2 \epsilon \right) \ .
\end{equation}
Furthermore, we define an analogous set of parameters for the kinetic
function $Z(\phi)$
\begin{equation}
\label{lambda}
\lambda\equiv M_p \frac{Z'}{Z} \ ,
\end{equation}
\begin{equation}
\label{kappa}
\kappa\equiv M_p^2 \frac{Z''}{Z} \ .
\end{equation}
The spectral index is then given by
\begin{equation}
n - 1 = \frac{1}{Z} \left( 2 \eta - 6 \epsilon - \sqrt{2 \epsilon} \lambda 
\right) +\dots \ .
\end{equation}
Here, we drop all terms
higher order in the slow-roll parameters.  Note that, in the absence of
cancellations, smallness of $n-1$ implies
\begin{equation}
|\sqrt{2 \epsilon} \lambda| \ll Z \ .
\label{indexconstraint}
\end{equation}

Finally, the running spectral index is
\begin{equation}
\frac{dn}{d \ln k} 
=   \frac{1}{Z^2} \left[
\left( -2 \xi + 16 \epsilon \eta - 24 \epsilon^2 \right)
+ 2 \epsilon \kappa - 4 \epsilon \lambda^2
+ \sqrt{2 \epsilon} \lambda \left( 3 \eta - 8 \epsilon \right) 
\right] + \dots \ .
\label{eq:runningslownoncanon}
\end{equation}
Again, we neglect terms higher order in
the slow-roll parameters. We also assume that the counting
of the number of derivatives gives an estimate of the order of the
parameters. This need not be the case, and some counterexamples
have been found \cite{DodelsonStewart}.

\subsection{k-essence and Tachyon-like Actions}
For completeness, let us mention another form of $p(X,\phi)$; the so-called k-essence 
form~\cite{Armendariz-Picon:2000dh,Armendariz-Picon:2000ah}
\begin{equation}
p(X,\phi) = \widetilde{p} (X) V(\phi) \ .
\end{equation}
The tachyon action considered in \cite{Sen}
is also of this form. The pressure and energy density are
\begin{eqnarray}
p &=& V(\phi) \widetilde{p} (X) \nonumber \\
{\cal E} &=& V(\phi) \widetilde{\cal E} (X)
\end{eqnarray}
where $\widetilde{\cal E} \equiv 2 X \widetilde{p}_{,X} - \widetilde{p}
(X)$.  Clearly, the function $V(\phi)$ does not enter the expressions for $n-1$
or $dn/d \ln k$. Therefore, in such models, the constraint from large running becomes
a criterion to be satisfied by the form of the ``kinetic term''
$\widetilde{p} (X)$.

\subsection{Constraints}
For inflationary predictions, we will in this paper aim for the range
of values provided by the WMAP analysis~\cite{Peiris:2003ff}.  
(All error bars correspond to 1$\sigma$ error
bars.) These are
\begin{equation}
P^\zeta (k_0 = 0.002 \mbox{Mpc}^{-1}) = 8 \pi ( 2.95\times 10^{-9})(0.77
\pm 0.07) \ ,
\label{eq:powerobserve}
\end{equation}
\begin{equation}
n(k_0=0.002 \mbox{Mpc}^{-1}) =  1.10^{+0.07}_{-0.06} \ ,
\label{eq:nsmin1obs}
\end{equation}
\begin{equation}
\frac{dn}{d\ln k} = -0.042^{+0.021}_{-0.020} \ ,
\label{eq:largerunningobs}
\end{equation}
\begin{equation}
r(k_0=0.002 \mbox{Mpc}^{-1})  < 0.71 \ .
\end{equation}
This final value significantly constrains the usual slow-roll parameter
$\epsilon$.  The data represents a combined fit to nearly all CMB
data, large scale structure measurements from the 2dF survey and power spectrum data
on the scale of the Lyman $\alpha$ forest (see \cite{Peiris:2003ff} and
\cite{Spergel:2003cb} for data definitions and more details of the
analysis).  Except where noted, in the analysis below we may ignore
the constraint~(\ref{eq:powerobserve}) on $P^{\zeta}(k)$ since, 
at least classically, we are free
to adjust the height of the inflaton potential.

The number of efoldings before the end of
inflation at which a perturbation mode left the horizon is
\begin{equation}
N(k)\approx 60.4+\frac{2}{3}\ln [g_{*}(t_{RH})]+\frac{1}{3}\ln \left(\frac{T_{RH}}{V_{e}^{1/4}}\right)+\ln \left(\frac{V_{e}^{1/4}}{10^{16}\textrm{GeV }}\right)+\ln \left(\frac{H_{0}/h}{k/a_{0}}\right) \ ,
\end{equation}
where $V_{e}$ is a fiducial value of the inflaton potential at the end
of inflation and $g_*(t_{RH})$ is the number of effectively massless
degrees of freedom at the reheating temperature $T_{RH}$. Setting
$k/a_{0}\approx H_{0}$ corresponds to a minimum number of efoldings
$N_{\rm min}$ typically between 50 and 60, although with some
dependence on $T_{RH}$.  In most of our analysis we will simply take
$N(k=0.002 \mbox{Mpc}^{-1})$ to be somewhere between 50 and 60 without
worrying about the details of reheating.

Note that $dn/d\ln k\approx -dn/dN$, so that the magnitude of $dn/d\ln
k$ decreases with increasing $N$.  This means that, in general,
the magnitude of
$dn/d\ln k$ is increased by minimizing $N_{\rm min}$, which is
achieved by low scale inflationary models. Hence, lowering the
reheating temperature and the scale of the inflaton potential
generically lead to stronger running of the spectral index. For
example, if $V_{e}^{1/4}\sim 1\textrm{GeV }$ and $T_{RH}\sim 1$ MeV
then we require
\begin{equation}
N\approx 23 \ ,
\end{equation}
which contributes a factor $1/23^{2}\sim 0.002$ to $dn/d\ln k$, rather
than $1/50^{2}\sim 0.0004$ in the typical high-scale models. However,
it is difficult to achieve a successful inflation scenario at such a
low energy scale (see, for example~\cite{Lyth:1998xn}).

\subsection{Negative Running and the Requirement of Large Third Derivative}
In order to achieve negative running
of the order of the central value of \eqr{eq:largerunningobs}, the
running must be dominated by the $\xi$-term (the third derivative
term) of \eqr{eq:xidefine} in the canonically normalized inflaton
basis.

To see this, suppose $dn/d\ln k$ is dominated by terms other than the $\xi$-term.  
Then, using~(\ref{eq:nsmin1slowcanon}) and~(\ref{eq:runningslowcanon}), we obtain
\begin{equation}
\frac{d n}{d \ln k} \approx 24 \epsilon^2 + 8 \epsilon (n-1) \ .
\end{equation}
For this to be sufficiently negative, we must have
\begin{equation}
- (n - 1) > 3 \epsilon + \frac{0.004}{\epsilon} \ .
\label{eq:impossibineq}
\end{equation}
Since the right hand side of this inequality is minimized at
$\epsilon=0.035$, \eqr{eq:impossibineq} forces $(n -1) < - 0.2$,
which is ruled out at around the 5$\sigma$ level by \eqr{eq:nsmin1obs}.
Hence, in order to attain the requisite running the $\xi$ term must
dominate.

In the case of a nonminimal kinetic term (but still with two
derivatives), we appear to have extra freedom to adjust $dn/d\ln k$ by changing
$\lambda$ and $\kappa$ in \eqr{eq:runningslownoncanon}. However,  this freedom
in adjusting the cancellation should just
correspond to adjusting the third derivative term $\xi$ after
canonically normalizing.  The nonminimal kinetic models hence must be
seen as a convenient way obtaining a large third derivative in those
situations in which the field redefinition to a canonical basis is
possible.

\section{\label{sec:canonrecipe} Singular Method}
We saw in the last section that to obtain a large \( dn/d\ln k
\), one must maximize the third derivative \( V'''/V \) while
minimizing \( V'/V \) and \( V''/V \). In this section we describe a
recipe for constructing an inflaton potential with these properties.

We begin by considering a singular
limit of what is required, one in which \( V'''/V \) diverges while \( V'/V \) and \( V''/V \)
remain regular. It is convenient to define a new function $f(\phi)$ by
\begin{equation}
V=V_{0}e^{f(\phi )} \ ,
\end{equation}
in terms of which
\begin{equation}
n-1=-f'^{2}+2f'' \ ,
\end{equation}
and
\begin{equation}
\frac{dn}{d\ln k}=2f'(f'f''-f''') \ ,
\end{equation}
where \( \phi  \) is evaluated at around \( 60 \) e-foldings before
the end of inflation, corresponding to the field value \( \phi _{*} \).
We require \( f' \) and \( f'' \)
to be regular at $\phi=\phi_*$, but $f'f'''$ to be singular there.
Note that it is \emph{insufficient} merely to choose
a function for which \( f'''(\phi _{*}) \) diverges, since the product
\( f'f''' \) may be regular even though \( f''' \) is irregular. 

To construct a suitable \( f \) define $K(\phi)$ by
\begin{equation}
\label{eq:kinf}
f(\phi )=s\int ^{\phi }\sqrt{K(x)}dx \ ,
\end{equation}
where \( s=\pm 1 \) is a sign. Our condition on \( f \) then
implies that \( K(\phi ) \) must satisfy
\begin{equation}
\label{eq:property1}
K(\phi_*)\ \ \ \ \textrm{ is regular}
\end{equation}
\begin{equation}
\label{eq:keycond}
\left.\frac{K'}{\sqrt{K}}\right|_{\phi_*} = 2 \left.\frac{d}{d\phi} \sqrt{K}\right|_{\phi_*} \ \ \ \ \textrm{ is regular}
\end{equation}
\begin{equation}
\label{eq:property3}
K''(\phi_*)\ \ \ \ \textrm{ is singular} \ .
\end{equation}
Therefore we require a \( C^{1} \) function \( K(\phi) \), for which \( K'(\phi) \) is discontinuous
at $\phi_*$ and $K|_{\phi \sim \phi _{*}}\neq 0$.

Although we cannot rule out the possibility of
\( K|_{\phi \sim \phi _{*}}=0 \), it is difficult to satisfy Eq.
(\ref{eq:keycond}) in such cases. The slow-roll parameters can be
written in terms of \( K \) as
\begin{equation}
\epsilon =\frac{K}{2} \ ,
\end{equation}
\begin{equation}
\eta =K+\frac{sK'}{2\sqrt{K}} \ ,
\end{equation}
and the corresponding observables are
\begin{equation}
n-1=-K+\frac{sK'}{\sqrt{K}}
\end{equation}
\begin{equation}
\frac{dn}{d\ln k}=s\left(\frac{K'}{\sqrt{K}}\right)K+\frac{1}{2}\left(\frac{K'}{\sqrt{K}}\right)^{2}-K''.
\end{equation}
Note that the observables take on a much simpler form in terms of \( K \)
compared to the expression in terms of \( f \). Additionally, these expressions
do not contain large numbers compared to when they are
expressed in terms of \( \epsilon  \) and \( \eta  \). One of the main challenges 
in obtaining strong running is now clear;
because \( \epsilon  \) is small, we must choose \( K \) to be
small, but choosing $K$ too small unacceptably increases \( n-1 \). 

Further, in order to have inflation at all we must ensure that slow-roll is valid 
throughout inflation. In other words, the direction in which \( \phi  \)
rolls is the same direction in which \( \epsilon  \) increases. This
implies that 
\begin{equation}
\textrm{ sign}\left[\left.\frac{d\epsilon }{d\phi }\right|_{\phi _{*}}\right]=-\textrm{sign}[V'(\phi _{*})] \ ,
\end{equation}
although, strictly speaking, this condition need not hold if there
are unusual features (``bumps") in the potential. This translates into the condition
\begin{equation}
\textrm{sign}[K']=-s \ ,
\end{equation}
which essentially fixes the sign \( s \).

Having dealt with an idealized singular limit, recall now that
we do not want \( dn/d\ln k \) to actually diverge at $\phi_*$. We therefore
regularize (smooth out) \( K'' \). One may accomplish
this either by adding small terms to remove the singularity, or by arranging
that the inflaton never quite reaches \( \phi _{*} \). 

Let us now restate the findings of this section as a simple recipe.
\begin{enumerate}
\item Choose a real differentiable function \( K(\phi ) \) which, at some field value $\phi_*$,
has $K$ and $d\sqrt{K}/d\phi$ small but nonzero and continuous, but $K''$ diverging 
negatively.
\item Define \( \tilde{K}(\phi ) \) as either a smoothed out version of \(
K(\phi ) \) (such that the singularity in \( K'' \) appears in 
\( \tilde{K} \) only in the limit that some new
``smoothing'' parameter vanishes: i.e. if $\lambda$ is a parameter
introduced for the purpose of regularizing, \( \lim _{\lambda
\rightarrow 0}\tilde{K}(\phi _{*})=K(\phi _{*}) \)) or as the original \(
K(\phi ) \) itself if \( \phi \) never reaches \( \phi _{*} \) during
inflation.
\item Define the inflaton potential via
\begin{equation}
V=V_{0}\exp\left(s\int^{\phi }\sqrt{\tilde{K}(x)}dx\right) \ ,
\end{equation}
where the sign \( s \) is chosen by \(
s=-\textrm{sgn}[\tilde{K}'(\phi _{e})] \) where \( \phi _{e} \) is the
end of the inflation determined by \( \tilde{K}(\phi _{e})=2 \).
\end{enumerate}

A consequence of this analysis is that we cannot choose \( K \) to be a monomial 
since then it would vanish at the singular point. This is why the example of
\cite{Kosowsky:1995aa}, which we consider in appendix 
\ref{kosowskyturner}, and for which the potential is of the form $ V_0
\exp(-\alpha \phi^b)$), does not result in a sufficiently large
$dn/d\ln k$.

\section{\label{sec:canonexamples}Singular Method Examples}

\subsection{Simple \protect\( K\protect \) but complicated \protect\( V\protect \)}
We begin with a simple example, choosing 
\begin{equation}
K=K_{0}+g\phi ^{\alpha } \ ,
\label{eq:knaive1}
\end{equation}
where \( K_{0} \), \( g \), and \( \alpha  \) are nonvanishing
constants. This implies
\begin{equation}
K''=g\alpha (\alpha -1)\phi ^{\alpha -2} \ ,
\label{eq:knaive2}
\end{equation}
which is singular at \( \phi _{*}=0 \) if \( \alpha <2. \) As required,
\( K \) itself does not vanish at \( \phi =\phi _{*} \), but the magnitude
of \( K \) can be small near \( \phi =\phi _{*} \) if \( K_{0} \)
is small. The magnitude of 
\begin{equation}
\frac{d}{d\phi}\left(\sqrt{K}\right)=\frac{g\alpha \phi ^{\alpha -1}}{2\sqrt{K_{0}+g\phi
^{\alpha }}}
\label{eq:knaive3}
\end{equation}
is also small, as required, if \( \alpha >1 \). Finally,
we want to make sure \( K'' \) does not truly diverge as \( \phi \rightarrow \phi _{*} \).
We can accomplish this by introducing a term \( m^{2} \) giving
\begin{equation}
\tilde{K}=K_{0}+g(\phi ^{2}+m^{2})^{\alpha /2} \ ,
\end{equation}
\begin{equation}
\tilde{K}''=g\alpha (m^{2}+\phi ^{2})^{\frac{\alpha -4}{2}}[m^{2}+\phi ^{2}(\alpha -1)] \ ,
\end{equation}
\begin{equation}
\frac{d}{d\phi}\left(\sqrt{\tilde{K}}\right)=\frac{g\alpha \phi (m^{2}+\phi ^{2})^{\frac{\alpha }{2}-1}}{2\sqrt{K_{0}+g(m^{2}+\phi ^{2})^{\alpha /2}}} \ ,
\end{equation}
which can then be compared with with Eqs.~(\ref{eq:knaive1}),
(\ref{eq:knaive2}), and (\ref{eq:knaive3}).

The resulting potential is 
\begin{equation}
V(\phi )=V_{0}\exp \left[ s\int _{0}^{\phi }dx\sqrt{K_{0}+g(x^{2}+m^{2})^{\alpha }}\right] \ ,
\end{equation}
where \( s=-\textrm{sign}[\tilde{K}'(\phi _{e})] \), (so that 
\( s=-\textrm{sign}[g] \) if \( \phi >0 \)). We choose $s=-1$
(corresponding to \( g>0 \)) so that the field starts near the
origin and rolls away from the origin (since \( \phi _{*}=0 \)).
The field value at the end of inflation, determined by \( \epsilon (\phi _{e})=\frac{1}{2}[K_{0}+g(m^{2}+\phi _{e}^{2})^{\alpha /2}]=1 \), is then
\begin{equation}
\phi _{e}= \sqrt{\left( \frac{2-K_{0}}{g}\right) ^{2/\alpha }-m^{2}} \ .
\label{eq:phieend}
\end{equation}
The observable parameters are then
\begin{equation}
n-1=-K_{0}-g(m^{2}+\phi ^{2})^{\alpha /2}+\frac{s\alpha g\phi (m^{2}+\phi ^{2})^{(\alpha -2)/2}}{\sqrt{K_{0}+g(m^{2}+\phi ^{2})^{\alpha /2}}}
\end{equation}
and
\begin{eqnarray}
\frac{dn}{d\ln k} & = & \frac{1}{2}g\alpha (m^{2}+\phi
 ^{2})^{\frac{\alpha -4}{2}} \left[ -2(m^{2}+(\alpha -1)\phi ^{2})
 +\frac{\alpha g\phi ^{2}(m^{2}+\phi ^{2})^{\alpha /2}}{K_{0}
 +g(m^{2}+\phi ^{2})^{\alpha /2}} \right. \nonumber \\ 
 & & \left. +2s\phi
 (m^{2}+\phi ^{2})\sqrt{K_{0}+g(m^{2}+\phi ^{2})^{\alpha /2}}\right] \ .
\end{eqnarray}

Since it is not possible to integrate analytically for \( \phi
_{N_{*}} \), we give numerical results to demonstrate that we can get
the desired \( n-1 \) and running \( dn/d\ln k \). We choose
\( \alpha =3/2 \) and \( K_{0}\approx 10^{-3} \), which yields small $\epsilon$.
Requiring that the running occur about 60 e-folds before the end of
inflation implies that we cannot make \( g \) too small, and we choose \( g=0.015 \). 
This value of \( g \) corresponds to \(
N\approx 55 \). Finally, we tune \( m^{2}\approx 0.01 \) to obtain
\begin{equation} \epsilon (\phi\approx 0)\approx 7\times
10^{-4}
\end{equation}
\begin{equation}
(n-1)|_{\phi\approx 0}\approx -1.5\times 10^{-3}
\end{equation}
\begin{equation}
\left.\frac{dn}{d\ln k}\right|_{\phi \approx 0}\approx -0.07
\end{equation}
which is the desired result.  Here $\phi\approx 0$ simply means
$\phi^2 \ll 10^{-3}$. Note that, instead of tuning \( m^{2}\approx 0.01 \),
we could have tuned the initial condition for \( \phi  \)
(starting slightly away from \( 0 \)), after setting \( m^{2}=0 \).
Note also that, to have the desired inflationary history, the inflaton must
begin rolling very close to the origin. It may
be possible, for example, to use thermal effects to place the inflaton at this position. 
Finally, note that the smoothing of \( K \) through the \( m^{2} \)
term yields a potential that is generically well
defined, even for negative values of \( \phi  \).

\subsection{A simpler \protect\( V\protect \)}
The potential in the previous example turned out to be complicated
because the integral of \( \sqrt{\tilde{K}} \) did not have an
analytic expression. Here we choose \( \tilde{K} \) to obtain a
simpler looking potential.  Choose
\begin{equation} K=(K_{0}+g\phi
^{\alpha })^{2} \ ,
\end{equation}
which has 
\begin{equation}
K''=2\alpha g\phi ^{\alpha -2}[(\alpha -1)K_{0}+(2\alpha -1)g\phi ^{\alpha }]
\end{equation}
\begin{equation}
\frac{d}{d\phi }(\sqrt{K})=\alpha g\phi ^{\alpha -1} \ .
\end{equation}
Again, we must choose \( 1<\alpha <2 \) so that \( K \) behaves appropriately
near the critical point \( \phi =\phi _{*}=0 \).
To maintain the simplicity of the potential, we set  $\tilde{K}=K$
and assume that \( \phi  \) never reaches the singular point \( \phi =0 \)
(since it is rolling away from the origin during inflation). This
leads to the potential
\begin{equation}
\label{eq:potentialsimp}
V(\phi )=V_{0}\exp\left[-\left(K_{0}\phi +\frac{g}{1+\alpha }\phi ^{1+\alpha }\right)\right] \ ,
\label{eq:unsmoothednonan}
\end{equation}
which is well defined for both positive and negative values of \( \phi  \)
if \( 1+\alpha =n/r \), where \( r \) is an odd integer and \( n \)
is an integer relatively prime to \( r \). This form of the potential
is simple, as promised. The end of inflation occurs at 
\begin{equation}
\phi_e =\left( \frac{\sqrt{2}-K_{0}}{g}\right) ^{1/\alpha }
\end{equation}
and the inflationary parameters are
\begin{equation}
\epsilon (\phi )=\frac{1}{2}(K_{0}+g\phi ^{\alpha })^{2}
\end{equation}
\begin{equation}
n-1=-2\alpha g\phi ^{\alpha -1}-(K_{0}+g\phi ^{\alpha })^{2}
\end{equation}
 \begin{equation}
\frac{dn}{d\ln k}=-2\alpha g\phi ^{\alpha -2}(K_{0}+g\phi ^{\alpha })(\alpha +K_{0}\phi +g\phi ^{\alpha +1}-1) \ .
\end{equation}
Again, for illustrative purposes, we choose
\( \alpha =5/3 \) and, since \( \epsilon \) should not be big, we choose
\( K_{0}=10^{-3} \). Finally \( g \) is tuned to give the desired
value of \( dn/d\ln k=-0.04 \) at 60 efoldings (\( \phi =
\phi_{60}=  5\times 10^{-6} \) ) by setting \( g=0.317 \).  The
resulting inflationary predicitions can be written as
\begin{equation}
\epsilon(\phi_{60})= 5 \times 10^{-7}
\end{equation}
\begin{equation}
(n-1)|_{\phi_N} = -3 \times 10^{-4}
\end{equation}
\begin{equation}
\frac{dn}{d\ln k} = -0.04 \ .
\end{equation} 
Although we have achieved large running, this example, like the previous one, still suffers from
the fact that the spectral index is always negative instead of running
from positive to negative.

\section{\label{sec:normalization}Non-canonical Kinetic Terms and The Connection to Field Redefinitions}
Consider again the special case of our general Lagrangian given by ${\cal L}=Z(\phi)X-{\tilde V}$.
The field redefinition that brings the non-minimal kinetic term into a canonical basis is
\begin{equation}
\Phi(\phi) =\int ^{\phi }\sqrt{Z(x)}dx \ ,
\end{equation}
for \( \Phi >0 \). For any given choice of \( \tilde{V}(\phi ) \)
one may compute \( \sqrt{Z(\phi )} \)
through the equation
\begin{equation}
\ln \left[\frac{\tilde{V}(\phi )}{V_{0}}\right]=s\int ^{\Phi }\sqrt{\tilde{K}(x)}dx \ ,
\end{equation}
or, slightly more explicitly,
\begin{equation}
\sqrt{Z(\phi )}=\frac{\tilde{V}'(\phi )}{\tilde{V}(\phi )}\frac{1}{s\sqrt{\tilde{K}(\Phi (\phi ))}} \ .
\end{equation}
Consider the simple potential given by Eq.~(\ref{eq:potentialsimp})
with \( \alpha =5/3 \). We can choose 
\begin{equation}
\label{eq:analpot}
\tilde{V}(\phi )=V_{0}\exp \left[-\left(K_{0}\phi ^{3}+\frac{3g}{8}\phi ^{8}\right)\right] \ ,
\end{equation}
which gives
\begin{equation}
\sqrt{Z(\phi )}=3\phi ^{2} \ .
\end{equation}
Here one sees the important role that may be played by the nonminimal
kinetic term: even though the required potential structure of
\eqr{eq:unsmoothednonan} is nonanalytic, it can be obtained from an
analytic potential of the form \eqr{eq:analpot} due to the field
redefinition arising from a nonminimal kinetic
structure. Unfortunately, the potential of Eq.~(\ref{eq:analpot}) is
still hard to motivate from a short distance physics point of view,
although it is at least analytic.  In general, however, classifying
those models, with $Z(\phi)\neq 0$, is more difficult than in the
canonical case, since there are two free functions of $\phi$ that
enter the Lagrangian density.

\section{\label{sec:functional} Potential as a functional of spectral index}
For convenience, we define \( j\equiv \sqrt{\tilde{K}} \), in terms of which
\begin{equation}
\label{eq:indexeq}
n(\phi )-1\equiv I(\phi )=-j^{2}(\phi )+2sj'(\phi ) \ ,
\end{equation}
which yields
\begin{equation}
\frac{dn}{d\ln k}=-sj(\phi )\frac{dI(\phi )}{d\phi } \ .
\end{equation}

Expanding about \( n-1=0 \) by letting \( I\rightarrow \lambda I \),
where \( \lambda  \) is a book-keeping perturbation parameter, we
can write the solution in a perturbation series to second
order in \( \lambda \)
\begin{equation}
\label{eq:pertdef}
j(\phi )=j_{0}(\phi )+\lambda j_{1}(\phi )+\lambda ^{2}j_{2}(\phi ) \ ,
\end{equation}
where we take the \( \lambda \rightarrow 1 \) limit at the
end. This yields
\begin{equation}
j_{0}(\phi )=\frac{j_{0}(\phi _{i})}{1-\frac{j_{0}(\phi _{i})}{2s}(\phi -\phi _{i})} \ ,
\end{equation}
\begin{equation}
\label{eq:j1int}
j_{1}(\phi )=\frac{1}{\left[ 1-\frac{j_{0}(\phi _{i})}{2s}(\phi -\phi _{i})\right] ^{2}}\int _{\phi _{i}}^{\phi }dy\frac{I(y)}{2s}\left[ 1-\frac{j_{0}(\phi _{i})}{2s}(y-\phi _{i})\right] ^{2} \ ,
\end{equation}
\begin{equation}
j_{2}(\phi )=\frac{1}{\left[ 1-\frac{j_{0}(\phi _{i})}{2s}(\phi -\phi _{i})\right] ^{2}}\int _{\phi _{i}}^{\phi }dy\frac{j_{1}^{2}(y)}{2s}\left[ 1-\frac{j_{0}(\phi _{i})}{2s}(y-\phi _{i})\right] ^{2} \ ,
\end{equation}
where \( j_{0}(\phi _{i}) \) is an integration constant. In the
familiar cases, we require that the potential revert to a constant
in the limit that \( n-1 \) vanishes. Hence, many situations
will involve \( j_{0}(\phi _{i})=0 \), which implies \( j_{0}(\phi )=0 \).
Note that if \( j_{0}(\phi _{i})=0 \), then \( j(\phi _{i})=0 \)
to all orders in \( \lambda  \). In such cases, since
\( \epsilon =j^{2}/2 \), we should set \( \phi _{i} \) by the condition 
\( \epsilon (\phi _{i})=0 \). Since the number of efoldings
diverges when \( \epsilon =0 \), \( \phi _{i} \) should
generically be set outside of the inflationary field values if \( j_{0}(\phi _{i})=0 \).
However, as we will see below, in order for \( n-1 \) to change sign from blue
to red during inflation, \( \phi _{i} \) must be within the domain
of inflationary field values. Hence, for potentials of our interest,
we will generally require \( j_{0}(\phi _{i})\neq 0 \) \footnote{For
simple checks of our formalism, we will sometimes take $j_0(\phi_i)=0$.}.

Although the order of the perturbation seems to imply that \( j_{0}^{2}(\phi )\gg j_{1}(\phi ) \)
has been assumed, this is not true. In fact, we may explicitly check that,
when \( j_{0}(\phi _{i})=0 \), then \( j^{2}(\phi )\sim O(\lambda ^{2}) \)
becomes the perturbation term (instead of the source \( I \)) and
the perturbative solution can easily be verified to be the same as
above. The potential \( V(\phi ) \) obtained via Eq.
(\ref{eq:pertdef}) can be written as\begin{equation}
\label{eq:pertpotential}
V(\phi )=\frac{V_{0}}{\left( 1-\frac{j_{0}(\phi _{i})}{2s}(\phi -\phi _{i})\right) ^{2}}
\exp\left(s\left[\int ^{\phi }j_{1}(z)dz+\int ^{\phi }j_{2}(z)dz\right]\right) \ .
\end{equation}
As long as \( j_{2}\ll j_{1} \), we can neglect \( j_{2}(z) \) in
the analysis. On the other hand, if \( j_{2}(z)>j_{1}(z) \), even
though the perturbation approximation has broken down, the potential
may still qualitatively give the desired results, and hence, even in
such cases, it is worth checking the potential to see if the result is
useful. 

To gain intuition about this formalism, let us write down the formula
for the potential in the simplest case, in which \( j_{0}(\phi _{i})=0 \)
and \( j_{2}(\phi ) \) has been dropped. We obtain
\begin{equation}
\label{eq:simpeq}
V(\phi )\approx V_{0}\exp \left[\frac{1}{2}\int ^{\phi }dx\int _{\phi _{i}}^{x}dy\, I(y)\right] \ .
\end{equation}
Note that, at this level of approximation, (where only \( j_{1} \) has
been kept), the result is identical to using the approximation
\begin{equation}
n-1=2\eta -4\epsilon +O(2\epsilon ) \ ,
\end{equation}
which means that the approximation with $j_0(\phi_i)$ is strictly only 
valid in the limit \( \eta \gg \epsilon \).  In cases where we want
the spectrum to run from blue to red, we require that $\epsilon$ to
play some role to cancel against $\eta$.  Hence, for the most
interesting case, we should not set $j_0(\phi_i)=0$.  Let us now check
our formalism with some simple potentials.

\subsection{Monomial potential reconstruction}
Consider the monomial potential for which (see appendix)
\begin{equation}
\label{eq:nmin1exp}
I(\phi)=\frac{-b(2+b)}{\phi ^{2}}
\end{equation}
and\begin{equation}
\label{eq:epsasphi}
\epsilon =\frac{b^{2}}{2\phi ^{2}} \ .
\end{equation}
From Eq. (\ref{eq:nmin1exp}), we see that the asymptotic expansion
{}``parameter'' is 
\begin{equation}
\delta \equiv \frac{b(2+b)}{\phi ^{2}} \ .
\end{equation}
Although \( \delta  \) is a function and not a parameter, we expect
that, for those values of \( b \) and \( \phi  \) for which \( \delta \rightarrow 0 \),
the perturbative solution will be well approximated by an expansion in \( \delta  \). This implies that \begin{equation}
b=-1+\sqrt{1+\phi ^{2}\delta }
\end{equation}
which, when expanded about \( \delta =0 \), gives
\begin{equation}
\label{eq:banddelta}
b=\frac{1}{2}\phi ^{2}\delta +O(\delta ^{2}) \ .
\end{equation}
Hence, to leading order in \( \delta  \), we need only keep the leading
order $b$-dependence in the final potential. Additionally,
we make the simplifying assumption that the potential
becomes a constant in the \( b\rightarrow 0 \) limit (this
is consistent with Eq.(\ref{eq:epsasphi})). Hence, we set \begin{equation}
\label{eq:j0cond}
j_{0}(\phi _{i})=0 \ .
\end{equation}
Finally, we choose \( \phi _{i}=\infty  \), since \( \epsilon (\infty )=0 \).
Eq. (\ref{eq:j1int}) then gives
\begin{equation}
j_{1}(\phi )=\frac{b(2+b)}{2s\phi } \ .
\end{equation}
Integrating with respect to \( \phi  \) we find
\begin{equation}
V=V_{0}\phi ^{b} \ ,
\end{equation}
where we have used Eq. (\ref{eq:banddelta}) and kept the leading
\( \delta  \)-dependence in the exponent. Of course, strictly
speaking, our approximation breaks down when \( \delta \sim 1 \),
which generically occurs before the end of inflation, set by \( \epsilon =1 \).
Thus, the approximation is questionable whenever \( j_{2}(\phi )>j_{1}(\phi ) \), 
for which we can explicitly show
\begin{equation}
j_{2}(\phi )=\frac{-sb^{2}(2+b)^{2}}{8\phi } \ .
\end{equation}
However, the method allows us to at least make a systematic guess regarding 
the potential, motivated by the spectral index. Furthermore, it is important
to remember that we have made a convenient assumption 
( \( j_{0}(\phi _{i})=0 \)) to obtain this simple form of the potential.

\subsection{Dynamical Supersymmetry Breaking Motivated Potential}
In the model of Ref.~\cite{Kinney:1998dv} 
which is motivated by dynamical supersymmetry breaking 
(see appendix) the function obtained
for \( n-1 \) to leading order in $\alpha$ is
\begin{equation}
\label{eq:kinneyriottosource}
I(\phi )=\frac{2\alpha p(1+p)}{\phi ^{2+p}} \ ,
\end{equation}
where we have again chosen \( j_{0}(\phi _{i})=0 \) to yield a constant
potential in the \( \alpha \rightarrow 0 \) limit. Integrating
Eq. (\ref{eq:kinneyriottosource}), one finds
\begin{equation}
j_{1}(\phi )=-\frac{s\alpha p}{\phi ^{1+p}}\left[1-\left(\frac{\phi}{\phi _{i}}\right)^{1+p}\right] \ ,
\end{equation}
where we have again chosen \( \phi _{i}=\infty  \), since \( \epsilon (\infty )=0 \).
Note that, since \( j_{1}(\phi ) \) is an intrinsically negative quantity,
we have $s=-1$.
Using
\begin{equation}
\frac{dn}{d\ln k}\approx -sj_{1}(\phi )\frac{dI}{d\phi }
\end{equation}
we find
\begin{equation}
\frac{dn}{d\ln k}\approx \frac{-2\alpha ^{2}p^{2}(2+p)(1+p)}{\phi ^{2(2+p)}} \ ,
\end{equation}
which agrees with Eq. (\ref{eq:krdndlnk}). Finally, from Eq. (\ref{eq:simpeq}),
we obtain \begin{equation}
V=V_{0}\left[1+\frac{\alpha }{\phi ^{p}}\right]
\end{equation}
which also agrees with Eq. (\ref{eq:kinneyriottoorig}).

Thus far, we have not addressed how we would have known that
Eq. (\ref{eq:kinneyriottosource}) is the correct spectral index function
to use. The difficulty in general is not getting a large \( dn/d\ln k \)
at any particular time, but having 60 e-folds afterwards. What helps 
this model work is that \( dn/d\ln k \) is large at
the beginning of inflation and decreases during inflation. During 
inflation $\epsilon$ is decreasing and the potential has a negative slope given by
\begin{equation}
\textrm{sign}[sj_{1}(\phi )]=-1 \ ,
\end{equation}
allowing $\phi$ to reach larger field values.
More generically, using \( j_{1}\approx \frac{1}{2s}\int _{\phi _{i}}^{\phi }dyI(y)>0 \),
one can see that
the desired behavior of the slope of the potential comes simply from 
\begin{equation}
\label{eq:goodcondition}
I(\phi )>0 ~,
\end{equation}
assuming that \( \epsilon  \) is a monotonic function during inflation.
To summarize, having a blue spectrum naturally aids in attaining sufficient
inflation after the point at which \( dn/d\ln k \) is large.

\subsection{Wilson Line as an Inflaton}
In the extra dimensional model discussed in the appendix, the source
function \( I(\phi ) \) is given by Eq. (\ref{eq:nmin1renjie}).
We thus find
\begin{equation}
j_{1}(\phi )=\frac{1}{2sf_{\rm eff}^{2}}\left( q_{1}^{2}\left[ (\phi -\phi _{i})+4\frac{f_{\rm eff}}{q_{1}}\left(\cot \frac{q_{1}\phi }{2f_{\rm eff}}-\cot \frac{q_{1}\phi _{i}}{f_{\rm eff}}\right)\right] +I_{2}\right) \ ,
\end{equation}
where we have defined
\begin{equation}
I_{2}\equiv q_{2}^{2}\sigma \int _{\phi _{i}}^{\phi }dx\cos \frac{q_{2}x}{f_{\rm eff}}\csc ^{2}\frac{q_{1}x}{2f_{\rm eff}} \ ,
\end{equation}
which can be expressed in terms of hypergeometric
functions. 

Unfortunately, it is not very easy to reconstruct this potential using
our method.  On the other hand, since \( q_{2}/q_{1}\gg 1 \) and \(
\sigma \ll 1 \), this model provides the bumps that we discussed in
the introduction.  This model with bumps is special because instead of
the bump being at a special location, it is a periodic set of bumps,
relieving the special initial condition problem.

\section{\label{sec:bluetored} The Index Method}
To construct a potential of the form~(\ref{eq:pertpotential}), yielding a spectral index
that runs from blue to red, we must choose \( I(\phi ) \) to vanish
at a field value \( \phi =\phi _{x} \)
during inflaton. Moreover, if \( \phi _{*} \) occurs 60 e-folds before
the end of inflation, there must only be about 5 e-folds between \( \phi =\phi _{*} \)
and \( \phi =\phi _{x} \). This implies
\begin{equation}
\left| \int ^{\phi _{x}}_{\phi _{*}}\frac{d\phi }{\sqrt{2\epsilon (\phi )}}\right| \approx 5 \ ,
\end{equation}
while sufficient total inflation requires
\begin{equation}
\left| \int _{\phi _{x}}^{\phi _{e}}\frac{d\phi }{\sqrt{2\epsilon (\phi )}}\right| \approx 55 \ ,
\end{equation}
where \( \phi _{e} \) is the value of $\phi$ at the end of inflation. 
This means that generally\begin{equation}
\label{eq:asymmetry}
|\phi _{*}-\phi _{x}|\ll |\phi _{x}-\phi _{e}|
\end{equation}
is necessary. 

Another necessary condition is that \( \phi _{x} \) lie in the direction of the slow-roll. This
results in
\begin{equation}
\textrm{sign}[\phi -\phi _{x}]=\textrm{sign}\left[\frac{1}{2}\int _{\phi _{i}}^{\phi }dyI(y)\right] \ .
\end{equation}
Furthermore, since we demand \( n-1>0 \) near \( \phi =\phi _{*} \), an
explicit check reveals that the only possibilities are 
\begin{equation}
\phi _{*}<\phi _{i}\leq \phi _{x} 
\label{eq:onepossibility}
\end{equation}
or 
\begin{equation}
\phi _{x}<\phi_{i}\leq \phi _{*} \ ,
\label{eq:secondpossibility} 
\end{equation} with \( n(\phi _{i})-1\geq 0 \).  Since we do
not want many efolds before \( \phi \) reaches \( \phi _{x} \) and
since \( \phi _{i} \) is close to the location where \( \epsilon \) usually
reaches a minimum, we set \( \phi _{i}=\phi _{x} \) generically.  By
making this choice, we have made both the $j_1$ and $j_2$ contributions vanish
precisely where $n -1$ changes sign.  In
choosing \( j_{0}(\phi _{i}) \), we require
\begin{equation}
\label{eq:slowrollj0}
\frac{j^{2}_{0}(\phi _{i})}{2}\ll 1 \ ,
\end{equation}
since \( \epsilon (\phi _{i})=j^{2}_{0}(\phi _{i})/2 \) to all orders
in the perturbation \( \lambda  \).

To get a better sense of the requirement~(\ref{eq:asymmetry}),
let us parameterize \( \epsilon  \) as
\begin{equation}
\label{eq:eps1}
\epsilon \sim \left\{\begin{array}{ll}
\frac{c_{1}^{2}}{2}(\phi -\phi _{x})^{2n_{1}}+\frac{\Delta ^{2}}{2}
& \mbox{for $\phi <\phi _{x}<\phi _{e}$} \\
\frac{c_{2}^{2}}{2}(\phi -\phi _{x})^{2n_{2}}+\frac{\Delta ^{2}}{2}
& \mbox{for $\phi _{x}<\phi <\phi _{e}$} ~,
\end{array} \right.
\end{equation}
where $c_i$, $\Delta$, and $n_i >0$ are constants.
This yields 
\begin{eqnarray}
\int ^{\phi _{x}}_{\phi _{*}}\frac{d\phi }{\sqrt{2\epsilon (\phi )}} & \sim  & \theta \left[ (\phi _{x}-\phi _{*})-\left(\frac{\Delta }{c_{1}}\right)^{1/n_{1}}\right] \left(\frac{\Delta }{c_{1}}\right)^{1/n_{1}}\frac{1}{\Delta }\nonumber \\
 &  & +\theta \left[ \left(\frac{\Delta }{c_{1}}\right)^{1/n_{1}}-(\phi _{x}-\phi _{*})\right] \frac{(\phi _{x}-\phi _{*})}{\Delta }\label{eq:epsnearbeginning} 
\end{eqnarray}
and
\begin{eqnarray}
\int _{\phi _{x}}^{\phi _{e}}\frac{d\phi }{\sqrt{2\epsilon (\phi )}} & \sim  & \theta \left[ (\phi _{e}-\phi _{x})-\left(\frac{\Delta }{c_{2}}\right)^{1/n_{2}}\right] \left(\frac{\Delta }{c_{2}}\right)^{1/n_{2}}\frac{1}{\Delta }\nonumber \\
 &  & +\theta \left[ \left(\frac{\Delta }{c_{2}}\right)^{1/n_{2}}-(\phi _{e}-\phi _{x})\right] \frac{(\phi _{e}-\phi _{x})}{\Delta } \ ,
\end{eqnarray}
where \( \theta  \) is a step function with \( \theta (z)=1 \) for
\( z>0 \) and \( \theta (z)=0 \) for \( z<0 \) \footnote{Although for smooth functions, \( \epsilon  \)
does not have such a discontinuous jump at \( \phi =\phi _{x} \),
the functional behavior is supposed to be indicative of the average
long term behavior of the function rather than exact behavior.}. 
For sufficiently small $c_1$, such that
\begin{equation}
\label{eq:approxvalid1}
5c_{1}^{1/n_{1}}\ll \Delta ^{1/n_{1}-1} \ ,
\end{equation}
then we can generically realize
\begin{equation}
\label{eq:5efold}
\int ^{\phi _{x}}_{\phi _{*}}\frac{d\phi }{\sqrt{2\epsilon (\phi )}}\approx \frac{\phi _{x}-\phi _{*}}{\Delta }\approx 5 \ ,
\end{equation}
while if we require 
\begin{equation}
\label{eq:approxvalid2}
\phi _{e}>\phi _{x}+N_{tot}\Delta 
\end{equation}
we can achieve
\begin{equation}
\label{eq:55efolds}
\int _{\phi _{x}}^{\phi _{e}}\frac{d\phi }{\sqrt{2\epsilon (\phi )}}\approx (\frac{\Delta }{c_{2}})^{1/n_{2}}\frac{\alpha }{\Delta }\approx N_{tot} \ ,
\end{equation}
where \( \alpha \sim O(1) \) and \( N_{tot}\approx 55 \)
for a total of 60 e-foldings. Note that, in this setting, we see explicitly
the hierarchy 
\begin{equation}
\label{eq:required}
\phi _{e}-\phi _{x}>11(\phi _{x}-\phi _{*})
\end{equation}
as expected. It is important to note that Eq. (\ref{eq:required})
is a necessary but not sufficient condition to
achieve \( 60 \) e-foldings, since this also depends on the value of \( c_{2} \). Generically,
it is very hard to estimate Eq. (\ref{eq:55efolds}) accurately (hence
there is a large uncertainty \( \alpha  \) ) because the functional
form for \( \epsilon  \) over the entire duration of inflation can
be complicated. Furthermore, we must keep in mind that, in some cases, inflation
ends because \( \eta  \) becomes of order 1 before \( \epsilon  \). 

From Eq. (\ref{eq:goodcondition}), we have learned that having a
blue spectrum naturally helps one to obtain a large \( dn/d\ln k \)
because \( \epsilon  \) can decrease during inflation from the time
when \( dn/d\ln k \) is large and thus help inflation achieve sufficient
e-folds of expansion. Note that, although Eq. (\ref{eq:goodcondition})
was derived assuming that \( j_{0}(\phi _{i})=0 \), it is still a
good condition in general. More precisely, if \( j_{0}(\phi _{i})\neq 0 \),
then \( \epsilon  \) is decreasing during inflation if
\begin{equation}
\textrm{sign}[j_{0}(\phi )+j_{1}(\phi )]=\textrm{sign}\left[\frac{I(\phi )}{2}+\frac{j_{0}(\phi )^{2}}{2}+j_{1}(\phi )j_{0}(\phi )\right] \ .
\end{equation}
Since \( j_{0}+j_{1}>0 \), the desired
behavior of \( \epsilon  \) is generically attained if Eq.(\ref{eq:goodcondition})
is satisfied. As we will show more explicitly later, this
decrease in $\epsilon$ stops at a location near $\phi \approx \phi_i$
where $n -1$ approximately vanishes.

To recap, the general recipe for construction is as follows:

\begin{enumerate}
\item Write down \( I(\phi )\approx n(\phi )-1 \) which changes sign
at \( \phi =\phi _{i} \).
\item Compute the potential \( V \) using Eq. (\ref{eq:pertpotential}) with
\( j_{0} \) and \( j_{1} \).
\item Compute the slow-roll parameters using the exact first order slow
roll equations.
\item Choose the parameters introduced in \( I \) to satisfy the constraint
Eqs. (\ref{eq:slowrollj0}), (\ref{eq:5efold}),(\ref{eq:55efolds}),
and 
\begin{equation}
\label{eq:dndlnkgeneralconst}
\frac{dn}{d\ln k}(\phi =\phi _{i})\approx -0.02 \ ,
\end{equation}
where \( \Delta  \) and \( c_{i} \) are defined in Eq. (\ref{eq:eps1}). 
Note that \( c_{2} \) is generically difficult
to define, in which case, one must try to estimate the integral of
Eq. (\ref{eq:55efolds}) some other way. Also Eq. (\ref{eq:dndlnkgeneralconst})
can be replaced by \( dn/d\ln k \) evaluated at \( \phi _{*} \)
instead and the right hand side adjusted to achieve the desired
running. 
\item Check the validity of the approximations by ensuring
that Eq. (\ref{eq:approxvalid1}) and Eq. (\ref{eq:approxvalid2})
are satisfied.
\item Test the total number of efolds numerically.
\end{enumerate}
As we discuss in Sec.~\ref{sec:newconsistency}, the intriguing general
feature of this very wide class of models is that \( \epsilon \)
reaches its minimum  approximately where \( n-1 \) vanishes.

\subsection*{A Successful example}
We have made a number of attempts to produce a viable model using the method of
this section. 
For example, 
the simplest functional $I(\phi) = c (\phi - \phi_x)$ gives a spectrum that runs from blue to red 
but this simple model does not give sufficient efolds of inflation.
Although there are problems with the simplest attempts, successful examples 
can be found, and we present one here.
Consider a running index function of the form
\begin{equation}
I(\phi )=c\left(\frac{1}{\phi ^{6}}-\frac{1}{\phi _{i}^{6}}\right) \ .
\end{equation}
The potential generated by \( j_{0}+j_{1} \) is not particularly illuminating. 
However, expanding about \( \phi =\phi _{i} \) to second order,
we obtain a slightly simpler form
\begin{eqnarray}
V=\tilde{V}_{0}\left[ 1 \right. &+& sj_{0}(\phi _{i})\left( (\phi -\phi _{i})-\frac{c}{6\phi _{i}^{3}}\right) \nonumber \\
&+& \left. j_{0}(\phi _{i})^{2}\left( \frac{c^{2}}{72\phi _{i}^{6}}-\frac{c\phi _{i}^{3}(2\phi -3\phi _{i})}{12}+\frac{3}{4}(\phi -\phi _{i})^{2}\right) \right] \ ,
\end{eqnarray}
where \( \tilde{V}_{0}\equiv V_{0}e^{3c/(8\phi _{i}^{4})} \). Hence,
near \( \phi =\phi _{i} \), this is a particular type of hybrid inflationary
potential. 

Following our procedure, we now find the relevent slow parameters
to second order in \( (\phi -\phi _{i}) \)
\begin{equation}
\epsilon \approx \frac{j^{2}_{0}(\phi _{i})}{2}+\frac{-s3cj_{0}(\phi _{i})}{2\phi _{i}^{7}}(\phi -\phi _{i})^{2} \ ,
\end{equation}
\begin{equation}
n-1\approx \frac{-6c}{\phi _{i}^{7}}(\phi -\phi _{i})+\frac{21c}{\phi _{i}^{8}}(\phi -\phi _{i})^{2} \ ,
\end{equation}
\begin{eqnarray}
\frac{dn}{d\ln k}\approx \frac{6cj_{0}(\phi _{i})s}{\phi _{i}^{7}} &+& \frac{3cj_{0}(\phi _{i})(j_{0}(\phi _{i})\phi _{i}-14s)}{\phi _{i}^{8}}(\phi -\phi _{i}) \nonumber \\
&-& \frac{3c[3c+7j_{0}(\phi _{i})\phi _{i}^{5}(j_{0}(\phi _{i})\phi _{i}-8s)]}{\phi _{i}^{14}}(\phi -\phi _{i})^{2} \ .
\end{eqnarray}
These approximations are valid as long as
\begin{equation}
\label{eq:approxgood}
\left|\frac{\phi }{\phi _{i}}-1\right| \ll 1 \ .
\end{equation}
The behavior of \( \epsilon  \) for moderate \( \phi -\phi _{i} \),
neglecting corrections from \( j_{0}(\phi _{i}) \), is
\begin{equation}
\label{eq:epsforend}
\epsilon \approx \frac{c^{2}}{8\phi _{i}^{12}}\left(\phi -\frac{6}{5}\phi _{i}+\frac{\phi _{i}^{6}}{5\phi ^{5}}\right)^{2} \ ,
\end{equation}
which implies that the end of inflation is around
\begin{equation}
\phi _{e}\approx \pm\frac{2\sqrt{2}\phi _{i}^{6}}{c}+\frac{6}{5}\phi _{i} \ .
\end{equation}

Let us now choose the parameters for our potential. The parameters
of Eq. (\ref{eq:eps1}) are 
\begin{equation}
c_{1}=\sqrt{\frac{-s3cj_{0}(\phi _{i})}{\phi _{i}^{7}}} \ ,
\end{equation}
\begin{equation}
n_{1}=1,\textrm{ }\Delta =j_{0}(\phi _{i}),c_{2}=c_{1},n_{2}=1 \ ,
\end{equation}
\begin{equation}
\frac{dn}{d\ln k}\approx \frac{6cj_{0}(\phi _{i})s}{\phi _{i}^{7}} \ ,
\end{equation}
where we will see below that setting \( c_{2}=c_{1} \) is not a particularly
good approximation for finding the total number of e-folds (neither
is using Eq.(\ref{eq:epsforend}), although that equation happens
to be useful for finding the end of inflation). The
constraint equations then become
\begin{equation}
\frac{6cj_{0}(\phi _{i})s}{\phi _{i}^{7}}\approx -0.02 \ ,
\end{equation}
\begin{equation}
\frac{\alpha }{\sqrt{\frac{-sc3j_{0}(\phi _{i})}{\phi _{i}^{7}}}}\approx N_{tot} \ ,
\end{equation}
\begin{equation}
|\phi _{i}-\phi _{*}|\approx 5j_{0}(\phi _{i}) \ ,
\end{equation}
where we will see numerically later that \( \alpha  \) can be as
large as \( 6 \), meaning that \( \epsilon  \) increases much more
slowly than can be extrapolated from the behavior near \( \phi =\phi _{i} \).
This is, of course, what we expected by construction. The first two
of these equations can be rewritten as\begin{equation}
c\approx \frac{-0.017\phi _{i}^{7}}{s|\phi _{i}-\phi _{*}|} \ ,
\end{equation}
\begin{equation}
N_{tot}\approx 10\alpha \ .
\end{equation}
Eqs. (\ref{eq:approxvalid1}) and (\ref{eq:approxvalid2}) then require
\begin{equation}
5\sqrt{\frac{-s3cj_{0}(\phi _{i})}{\phi _{i}^{7}}}<1
\end{equation}
and 
\begin{equation}
N_{tot}j_{0}(\phi _{i})+\phi _{i}<\frac{6}{5}\phi _{i}+\frac{2\sqrt{2}}{c}\phi _{i}^{6} \ .
\end{equation}
Only the second of these conditions yields the nontrivial constraint
\begin{equation}
0<\frac{1}{5}\phi _{i}+\left(-848.5\frac{s}{\phi _{i}}-10\alpha\right)j_{0}(\phi _{i}) \ .
\end{equation}

\begin{figure}
{\centering \includegraphics{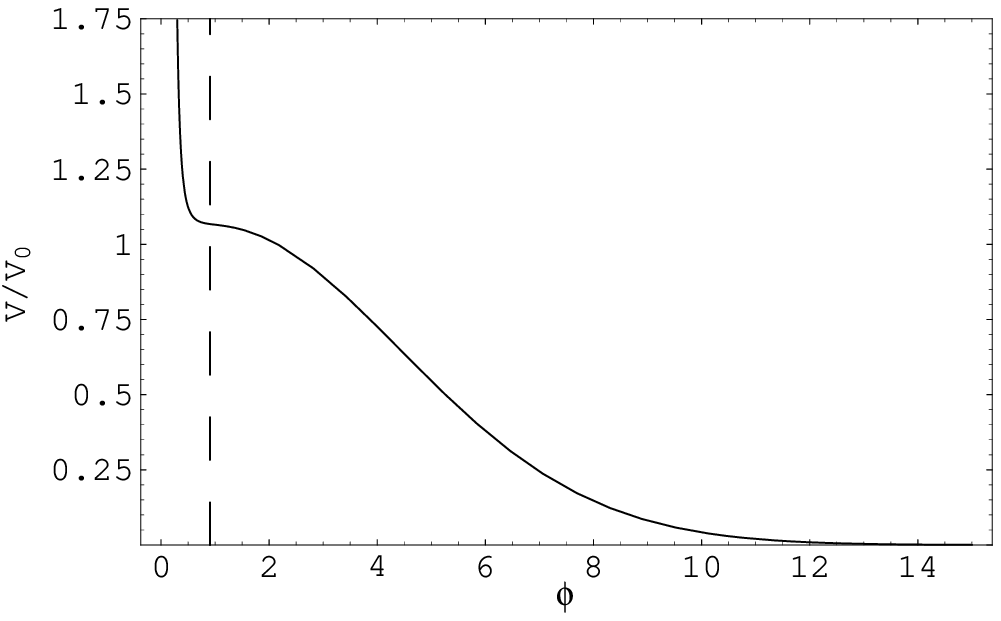} \par}
{\centering \includegraphics{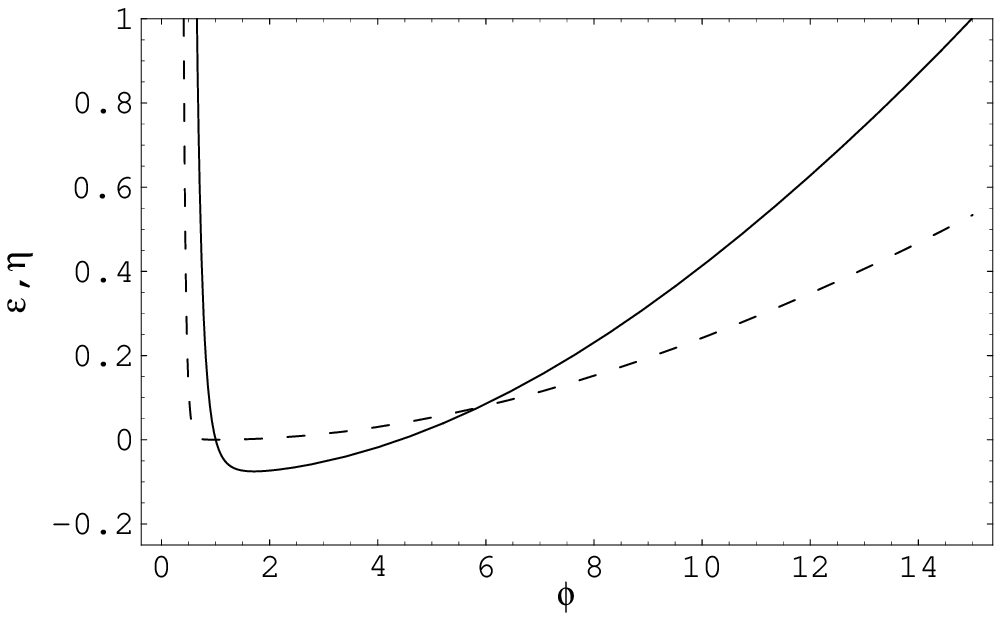} \par}
\caption{\label{fig:potentialandeta}The upper figure shows the potential as a function of the inflaton
field \protect\( \phi \protect \). The vertical long-dashed curve
in the upper figure corresponds to a field value \protect\( 63\protect \)
efolds before the end of inflation. In the lower figure, the solid
curve corresponds to the behavior of \protect\( \eta \protect \)
while the short-dashed curve corresponds to the behavior of \protect\( \epsilon \protect \).
Note that inflation ends due to \protect\( \eta \protect \)
and not \protect\( \epsilon \protect \). Also, one can clearly see
that \protect\( \epsilon \protect \) has a minimum near \protect\( \phi =1\protect \)
where \protect\( n-1\protect \) changes sign.
About 10 e-folds after the dashed line corresponds to $\phi\approx 1.1$.}
\end{figure}

\begin{figure}
{\centering \includegraphics{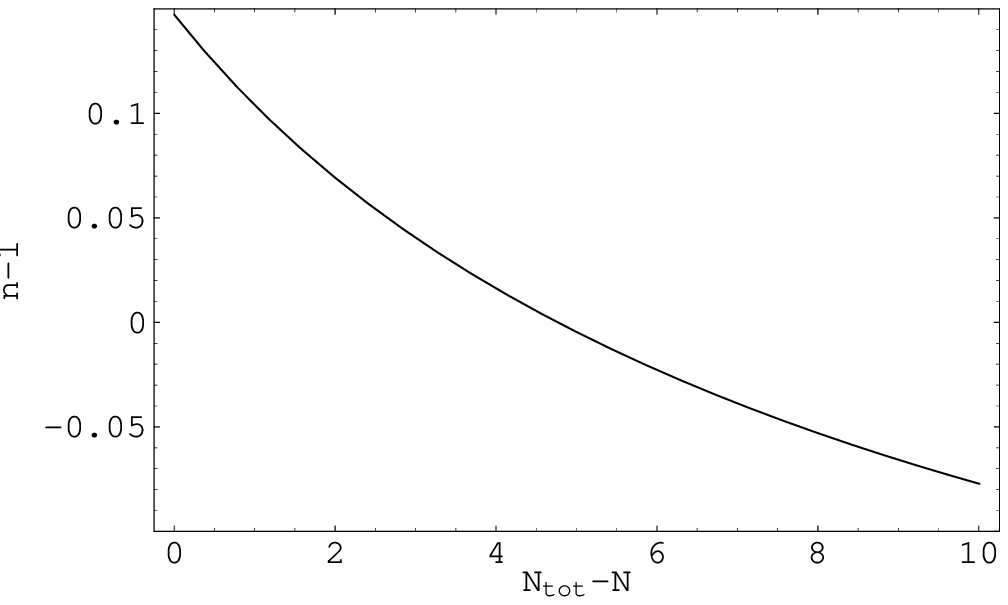} \par}
{\centering \includegraphics{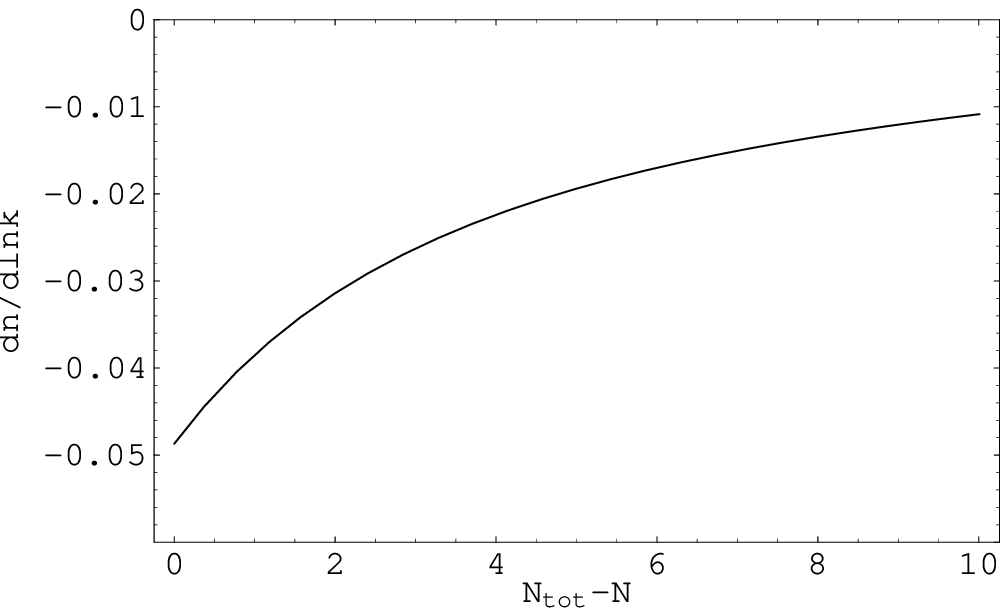} \par}
\caption{\label{fig:nanddndlnk} \protect\( n-1\protect \) and \protect\( dn/d\ln k\protect \) as
a function of \protect\( N_{tot}-N\protect \) where \protect\( N\protect \)
is the number of e-folds before the end of inflation and \protect\( N_{tot}\approx 63\protect \)
is the total number of efolds.}
\end{figure}

A viable example is then given by choosing \( \{\phi _{i}=1,|\phi
_{*}-\phi _{i}|=0.1\} \), for which our equations yield \(
\{s=-1,j_{0}(\phi _{i})=0.02,c\approx 0.167,\phi _{e}\approx 18\} \).
The potential, the spectral index, and its running for this model are
given in Figs.~\ref{fig:potentialandeta} and \ref{fig:nanddndlnk}.


\section{\label{sec:newconsistency} An Approximate Coincidence}
Here we derive the advertised result that there is an approximate
local minimum of $\epsilon$ at the length scale at which $n -1$
vanishes.  

At the most basic level, it is simple to see how such a consistency
condition can arise.  By elementary manipulation of the slow-roll
parameters, one can write
\begin{equation}
\frac{n - 1}{\sqrt{2 \epsilon}} + \sqrt{2 \epsilon} =
\pm\frac{\epsilon'(\phi)}{\epsilon}  \ ,
\label{eq:maincoincidence}
\end{equation}
where the upper sign is for $V'>0$ whereas the lower sign is for
$V'<0$.  This means that the minimum of $\epsilon$ is reached when
$n -1 \approx - 2 \epsilon$ which means that $n -1$ vanishes
slightly {\em before} the value at which $\epsilon$ reaches its minimum
$\phi=\phi_c$.  Linearizing $\epsilon$ and $\eta$ about $\phi_x$ where
$n(\phi_x) -1 = 0$, we can solve for $\phi_c$.  
We can then compute the number of efolds
$\Delta N$ that elapse between $\phi_c$ and $\phi_x$
\begin{equation}
\Delta N \approx \frac{ \epsilon(\phi_x) }{\xi(\phi_x) -10
\epsilon^2(\phi_x) }\approx \frac{ \epsilon(\phi_x) }{0.01 -10
\epsilon^2(\phi_x) } \ .
\end{equation}
Thus, if $\epsilon(\phi_x) <  0.01$, then $\Delta N < 1$
\footnote{Note that the precise value of $\phi_x$ where $n(\phi_x)-1
\equiv 0$ is dependent on higher order slow roll corrections, particularly
because the higher order parameter $\xi$ is typically forced to be at
least as large as $\epsilon$.  However, we have checked explicitly that
our conclusions in this section are not sensitive to that correction.
This is easy to understand given that the main arguments do not depend on
the precise details of the value of $\phi_x$, but merely the fact that
$n-1$ is typically much larger than $\epsilon$ except where $n-1=0$.  We
thank Jim Cline for bringing this issue to our attention.}.

From the perspective of the index formalism, the approximate minimum
for $\epsilon$ occurs near $\phi_x$ because of
Eqs.~(\ref{eq:onepossibility}) and (\ref{eq:secondpossibility}).
Namely, since $\epsilon \approx (j_0 + j_1+ j_2)^2$, we have the
$j_1$ and $j_2$ contributions vanishing precisely at $\phi_i$ which is
between the field value $\phi_*$ 
and the value
$\phi_x$.  Of course, one must note that, strictly speaking, $\phi_i$ is
not the value at which the derivative of $\epsilon$ vanishes.  The
effect is merely the same as saying that the $n -1$ term in
Eq.~(\ref{eq:maincoincidence}), which dominates in general, just happens
to be very small near $n -1=0$.

From a practical standpoint, it is not clear whether this can be
confirmed by measuring tensor perturbations.  The reason is simply
that this coincidence occurs most accurately for small $\epsilon$,
which in turn implies that the tensor to scalar amplitude ratio is
negligible.

\section{Summary and Conclusions}
Motivated by recent data, including that from the WMAP satellite, we
investigate the extent to which the scalar spectral index $n$ that
runs strongly negatively can be accomodated within inflationary
models.  We found that a running as large as the central value of
Eq.~(\ref{WMAP}) is difficult to achieve in common realizations of
slow-roll inflation as it requires a large third (field-)derivative
while maintaining small first and second derivatives.  Aside from the
fact that it is difficult to motivate potentials with these features
from a fundamental theory, it is not easy to sustain a large number of
efoldings (at least 60) for models with large $dn/d \ln k$.
Therefore, if current observations hold up, the large running of the
scalar spectral index could pose new challenges for inflationary model
building.

In view of these difficulties, we develop two methods to
systematically construct inflationary potentials with large running.
The first method, which we call the singular method, allows us to
generate arbitrarily large running without upsetting the slow-roll
requirements. However, this method by itself does not give rise to a
change in sign of $n-1$ within the range of the observable length
scales (about 5 efolds) nor does it give rise to a blue spectrum on
large scales -- features that are suggested by the recent cosmological
data including the WMAP data.  The second method, which we call the
index method, covers a broad range of slow-roll models that have a
blue spectrum on large scales, and is designed to construct models in
which $n-1$ runs from blue to red.  Obviously, these methods can be
combined together (as well as with other formalisms) to construct a
large variety of inflationary models.

We also uncover a fairly generic implication of strong negative 
running of the scalar spectral index $n$.  We show that in many
situations, there is an approximate local minimum of the slow-roll
parameter $\epsilon$ at the length scale at which $n -1$ vanishes.
This approximately implies that the strong running of the spectral
index requires a bump-like structure (defined by a region where the
slope of the potential reaches a local minimum) in the inflaton
potential.

At a more formal level, we also explore the extent to which the
constraints on the inflaton potential can be relaxed if the kinetic
term of the inflaton is non-canonical.  We found that we have extra
freedom in adjusting a large running while maintaining a small $n-1$
because of the new contributions from the field-dependent kinetic
term. Of course, this extra freedom simply corresponds to adjusting
the third derivative of the inflaton potential after we canonically
normalize the field.  However, such non-canonical kinetic terms may be
seen as a convenient way to generate inflaton potentials with unusually
large third derivatives starting from potentials that are more
physically motivated (e.g., potentials that are analytic functions of
the inflaton fields, etc).  Furthermore, non-minimal kinetic terms are
quite generic as they often appear in supersymmetric models, as well
as in string theory.  Therefore, cosmological data such as the running
spectral index may tell us something about the K\"{a}hler
potential. We hope to return to this and related issues in the future.

\acknowledgments 

We thank Jim Cline, Joanne Cohn, Scott Dodelson, Richard Easther,
Lisa Everett, Zoltan
Haiman, Edward Kolb, Julien Lesgourgues, Antonio Riotto, Uros Seljak,
Dominik Schwarz, and Liantao Wang for helpful discussions. We would
like to thank the Kavli Institute for Theoretical Physics, where this
work was begun, for kind hospitality and support. This work was
supported in part by the National Science Foundation (NSF) under grant
PHY-9907949. The work of GS is supported in part by funds from the
University of Wisconsin. The work of MT is supported in part by the
NSF under grant PHY-0094122, and is a Cottrell Scholar of Research
Corporation.

\section{Appendix: Models with Canonical Kinetic Terms}
\label{sec:regularmodels}
In this appendix, we analyze a sample of inflationary models to see
how large a value of $dn/d\ln k$ can be obtained, and at what price.
We apologize to many authors whose models we did not review due to
practicality constraints.  A much more extensive review of older
models can be found in \cite{Lyth:1998xn}.

\subsection{Monomials}
The best-known and most robust inflationary models are those exhibiting chaotic inflation. This may
be implemented with the simplest of potentials, namely monomials.

Consider a potential of the form
\begin{equation}
V(\phi )=V_{0}\left(\frac{\phi }{M_{p}}\right)^{b} \ ,
\end{equation}
where $b$ is a dimensionless parameter. The slow-roll parameters for this model are
\begin{eqnarray}
\epsilon &=& \frac{b^{2}M_{p}^{2}}{2\phi ^{2}} \ , \\
\eta &=& \frac{(b-1)bM^{2}}{\phi ^{2}} \ ,
\end{eqnarray}
and the third-derivative parameter is given by
\begin{equation}
\xi =\frac{(b-2)(b-1)b^{2}M_{p}^{4}}{\phi ^{4}} \ .
\end{equation}
In general in inflationary models, inflation ends when the first of the slow-roll conditions is violated. This
occurs at a field value $\phi_e$ defined by $\epsilon(\phi_e) = 1$. For the monomial potential this yields
\begin{equation}
\phi _{e}=\frac{bM_{p}}{\sqrt{2}} \ .
\end{equation}
Furthermore, the value of the field $\phi$ when there remain $N$ e-foldings before the end of inflation
is given by
\begin{equation}
\phi (N)=M_{p}\sqrt{\frac{b(b+4N)}{2}} \ .
\end{equation}

The quantities of primary interest in this paper, the scalar spectral index and its scale-dependence
are given by
\begin{eqnarray}
n-1 &=& \frac{-2(2+b)}{b+4N} \ , \\
\frac{dn}{d\ln k} &=& \frac{4}{(b+4N)}(n-1) \ .
\end{eqnarray}

Now, is it possible to obtain significant $dn/d\ln k$ in these models. The important point here is that
$N\geq 50$. Therefore, even for large values of $b$ we end up with a small $dn/d\ln k$. As an example, consider $b=20$, which gives $n-1=-0.2$, but a relatively small value $dn/d\ln k=-0.004$. Thus, only 
minor running of the spectral index is possible in minimal models with monomial potentials. For reference, 
note that the relative sizes of the terms contributing to $dn/d\ln k$ are
\begin{eqnarray}
-2\xi &=& -5.6\times 10^{-2} \ , \\
16\epsilon \eta &=& 2.5\times 10^{-1} \ , \\
-2\epsilon ^{2} &=& -2.0\times 10^{-1} \ ,
\end{eqnarray}
and so $\epsilon \eta$ and $\epsilon^{2}$ are both larger than the third derivative term.

\subsection{\label{kosowskyturner}
 Potentials with Powers in the Exponent}
We now turn to a model that was originally proposed~\cite{Kosowsky:1995aa} to obtain large running of the scalar index. Consider the following inflaton potential
\begin{equation}
V(\phi )=V_{0}e^{-\alpha \phi ^{b}} \ ,
\end{equation}
where $b$ is a dimensionless constant and $\alpha$ is a constant with dimensions of 
$[{\rm mass}]^{-b}$. The slow-roll and third-derivative parameters become
\begin{eqnarray}
\epsilon &=& \frac{\alpha ^{2}M_{p}^{2}b^{2}}{2}\phi ^{2(b-1)} \ , \\
\eta &=& \alpha M_{p}^{2}\phi ^{b-2}b[1+b(\alpha \phi ^{b}-1)] \ , \\
\xi &=& \alpha ^{2}M_{p}^{4}\phi ^{2(b-2)}b^{2}[2+3b(\alpha \phi ^{b}-1)+b^{2}(\alpha^{2}\phi ^{2b}-3\alpha \phi ^{b}+1)] \ .
\end{eqnarray}
As expected, the value of the scalar spectral index depends on the field value $\phi$, yielding
\begin{eqnarray}
n(\phi )-1 &=& -\alpha M_{p}^{2}\phi ^{b-2}b[b(\alpha \phi ^{b}+2)-2] \ , \\
\frac{dn}{d\ln k} &=& -2\alpha ^{2}M_{p}^{4}\phi ^{2(b-2)}(b-1)b^{2}(b-2+\alpha \phi ^{b}b) \ .
\end{eqnarray}

To complete the analysis it is important also to know that inflation ends at a field value
\begin{equation}
\phi _{e}=\left(\frac{\sqrt{2}}{\alpha M_{p}b}\right)^{\frac{1}{b-1}} \ ,
\end{equation}
and the number of efolds before the end of inflation at any given value
of $\phi <\phi _{e}$ is
\begin{equation}
N(\phi )=\frac{1}{b(2-b)M_{p}^{2}\alpha }(\phi _{e}^{2-b}-\phi ^{2-b}) \ ,
\end{equation}
where $\phi _{i}$ is the value of $\phi$ at the beginning of inflation. 

Using the above expressions we can rewrite the relevant quantities as
\begin{eqnarray}
n-1 &=& -\left[ \frac{2b(b-1)}{y}+b^{2}(\alpha M_{p}^{b})^{2-x}y^{-x}\right] \ , \\
\frac{dn}{d\ln k} &=& \left[ \frac{2b(b-1)}{y^{2}}+x\frac{b^{2}(\alpha M_{p})^{2-x}}{y^{1+x}}\right] b(2-b) \ ,
\end{eqnarray}
with
\begin{equation}
y\equiv \left[ \frac{2^{1-\frac{b}{2}}}{\alpha M_{p}^{b}b^{2-b}}\right] ^{\frac{1}{b-1}}-Nb(2-b) \ .
\end{equation}
By carefully tuning parameters one may obtain values of the inflationary observables that are close to
those that we seek. Our best case is shown in table~\ref{figtune1}.
\begin{table}[center]
\begin{tabular}{|c|c|}
\hline 
\( n-1 \)&
\( -0.185 \)\\
\hline 
\( \frac{dn}{d\ln k} \)&
\( -0.007 \)\\
\hline 
\( \epsilon  \)&
\( 0.065 \)\\
\hline 
\( \eta  \)&
\( 0.10 \)\\
\hline
\end{tabular}
\caption{\label{figtune1}Inflation characterization with the model choice
\protect\( \{b=2.01,\alpha =0.013\}\protect \).}
\end{table}
However, given this result, this type of potential is probably not favorable for the large values of $dn/d\ln k$ that may be required.

Once again, for reference, note that the relative sizes of the terms contributing to $dn/d\ln k$ are
\begin{eqnarray}
-2\xi &=& -1.2\times 10^{-2} \ , \\
16\epsilon \eta &=& 1.1\times 10^{-1} \ , \\
-2\epsilon ^{2} &=& -1.0\times 10^{-1} \ ,
\end{eqnarray}
and so $\epsilon \eta$ and $\epsilon^{2}$ are again both larger than the third derivative term
and hence an analysis with just the third derivative is inappropriate.

Since the above expressions leading to our conclusion are somewhat complicated, it is instructive
to consider the limit $\phi _{e}\gg \phi$ and to assume $b>2$. 
This then yields
\begin{equation}
N(\phi )\approx \frac{\phi ^{2-b}}{b(b-2)M_{p}^{2}\alpha } \ ,
\end{equation}
which, in this limit, allows us to express everything in terms of $N(\phi )$.
We obtain
\begin{equation}
n-1=\frac{-x}{N}-\frac{1}{(b-2)^{x}}(\alpha M^{b}b)^{2-x}\frac{1}{N^{x}} \ ,
\end{equation}
where
\begin{equation}
x\equiv \frac{2(b-1)}{b-2} \ .
\end{equation}
Furthermore,  since $\frac{dn}{d\ln k}\approx -\frac{dn}{dN}$, we have
\begin{equation}
\frac{dn}{d\ln k}\sim (n-1)/N \ .
\end{equation}
Since $|n-1|<0.1$ and $N\approx 50$, we can expect $\frac{dn}{d\ln k}\sim O(1)\times 10^{-3}$, with at
best $dn/d\ln k\sim 10^{-2}$, which is close to the numbers obtained from our rigorous analysis above.

\subsection{Running Mass Potentials}
Another class of potentials which may lead to large scale-dependence of the spectral index are the
so-called running mass potentials (see, for example~\cite{Lyth:1998xn,Covi:1998jp,Covi:1998mb,Covi:1998yr,Covi:2002th}). Consider 
\begin{equation}
V=V_{0}\left\{ 1-c\left[\ln\left(\frac{\phi }{\phi _{*}}\right)-\frac{1}{2}\right]\left(\frac{\phi }{M}\right)^{2}\right\} \ ,
\end{equation}
where the constant $V_{0}$ term dominates. In this model the slow-roll condition
\begin{equation}
\epsilon =\frac{8c^{2}M_{p}^{2}\phi ^{2}\ln ^{2}[\frac{\phi }{\phi _{*}}]}{[4M_{p}^{2}+c\phi ^{2}-2c\phi ^{2}\ln (\frac{\phi }{\phi _{*}})]^{2}}<1
\end{equation}
is satisfied if $c\phi ^{2}\ll M_{p}^{2}$. In this case,
we have
\begin{equation}
\epsilon \approx \frac{c^{2}}{2}\left(\frac{\phi }{M_{p}}\right)^{2}\ln ^{2}\left(\frac{\phi }{\phi _{*}}\right) \ .
\end{equation}
The other slow-roll parameter and the third-derivative parameter are
\begin{eqnarray}
\eta &=& \frac{-c\left[1+\ln \left(\frac{\phi }{\phi _{*}}\right)\right]}{1+c\frac{\phi ^{2}}{4M_{p}^{2}}-c\frac{\phi ^{2}}{2M_{p}^{2}}\ln \left(\frac{\phi }{\phi _{*}}\right)} \ , \\
\xi &=& \frac{-c\ln \left(\frac{\phi }{\phi _{*}}\right)}{\left[1+c\frac{\phi ^{2}}{4M_{p}^{2}}-c\frac{\phi ^{2}}{2M_{p}^{2}}\ln \left(\frac{\phi }{\phi _{*}}\right)\right]^{2}} \ .
\end{eqnarray}

Now consider the situation in which the end of inflation is controlled
by another field direction, as in hybrid inflation. In this case the condition for the end of inflation
is not $\epsilon \rightarrow 1$ while
rolling in the $\phi$ direction. Instead, one possibility is that inflation abruptly
ends when $\phi$ reaches a critical value $\phi _{c}$
(because of running off into another field direction)\footnote{Another
option is 
that the field evolves from small initial value to a large $\phi _{e}$
and that there is some stabilization mechanism such that $\phi$
does not run off to $\infty$. Of course, the logarithmic running
breaks down by that point.}. One can easily
carry out the required integration to obtain
\begin{equation}
\label{eq:Nrunningmass}
N(\phi )\approx -\frac{1}{c}\left\{\ln \left[\ln \left(\frac{\phi}{\phi _{*}}\right)\right]-\ln\left[\ln \left(\frac{\phi _{c}}{\phi _{*}}\right)\right]\right\} \ ,
\end{equation}
which yields
\begin{equation}
\phi =\phi _{*}\left(\frac{\phi _{c}}{\phi _{*}}\right)^{e^{-cN}} \ ,
\end{equation}
where the minus sign comes from the fact that $\ln (\phi /\phi _{*})<0$.
Clearly the double exponential sensitivity to $N$ seems to be the
key to obtaining a large running of the spectral index. 

After some algebra, the above expressions allow us to write
\begin{equation}
n-1 = -2c\left[1+e^{-cN}\ln \left(\frac{\phi _{c}}{\phi _{*}}\right)+{\cal O}\left(\frac{c\phi _{*}^{2}}{M^{2}(\phi _{c}/\phi _{*})^{2e^{-cN}}}\right)\right] \ ,
\end{equation}
and
\begin{equation}
\frac{dn}{d\ln k} \approx c(n-1+2c) \ .
\end{equation}

To obtain $dn/d\ln k \sim {\cal O}(10^{-2})$, we must choose $c\sim 10^{-1}$. 
Unfortunately, this gives $\exp (-cN)\approx 0.006$, which would give too small a suppression factor. 
To fight this suppression one must choose, for example, $\phi _{c}/\phi _{*}=10^{-23}$,
for which $e^{-50c}\ln (\phi _{c}/\phi _{*})=-0.36$. This logarithm 
is too large for the perturbative radiative ``correction'' to
be valid. 

If we nevertheless neglect the physics of the potential and
allow such a large logarithm, then, setting $\phi_*=M_p$,  Eq.~(\ref{eq:Nrunningmass}) implies
that a sufficient number (say 50) of efoldings can be obtained only if the initial value of $\phi$ is
approximately $0.7\phi _{*}$. This results in $n-1=-0.13$ and $dn/d\ln k=0.007$.

Of course, choosing $\phi _{c}$ even smaller results in a larger
$dn/d\ln k$, but the price is fine tuning the end of inflation
(controlled parametrically by $\phi _{c}$ and realistically by
the potential in the other field direction). For example, one obtains
$dn/d\ln k=0.018$ and $n-1=-0.017$ with $\phi _{c}=10^{-60}\phi _{*}$.

Once more, for reference, note that, in the case of $\phi _{c}=10^{-60}\phi _{*}$, 
the relative sizes of the terms contributing to $dn/d\ln k$ are
\begin{eqnarray}
-2\xi &=& 7\times 10^{-3} \ , \\
16\epsilon \eta &=& -3\times 10^{-4} \ , \\
-2\epsilon ^{2} &=& -2.33\times 10^{-6} \ ,
\end{eqnarray}
and so the third derivative term dominates in this example.

\subsection{Potentials Motivated by Dynamical Supersymmetry Breaking}
Ref.\cite{Kinney:1998dv} considers a potential of the form\begin{equation}
\label{eq:kinneyriottoorig}
V(\phi )=V_{0}(1+\frac{\alpha }{\phi ^{p}})
\end{equation}
which can be motivated from dynamical SUSY breaking. One can easily
compute\begin{equation}
\epsilon =\frac{\alpha ^{2}p^{2}}{2\phi ^{2}(\alpha +\phi ^{p})^{2}}
\end{equation}
\begin{equation}
n-1=\frac{\alpha p\{2[1+p]\phi ^{p}-\alpha (p-2)\}}{\phi ^{2}(\alpha +\phi ^{p})^{2}}
\end{equation}
\begin{equation}
\frac{dn}{d\ln k}=\frac{2\alpha ^{2}p^{2}\{(p-2)\alpha [\alpha +2(p+1)\phi ^{p}]-(1+p)(2+p)\phi ^{2p}\}}{\phi ^{4}(\alpha +\phi ^{p})^{4}}
\end{equation}
\begin{equation}
N=\textrm{sign}[\alpha ]\left\{ \frac{\phi _{e}^{2}-\phi ^{2}}{2p}+\frac{\phi _{e}^{p+2}-\phi ^{p+2}}{\alpha p(2+p)}\right\} 
\end{equation}
where \( \phi _{e} \) is the field value at the end of inflation.
If the inflationary scenario is to be hybrid, \( |\phi ^{p}|\gg |\alpha | \).
In that case, we must have \( \alpha >0 \) to have a blue spectrum.\footnote{This is obvious given that to have a blue spectrum, the curvature described by $\eta$ has to be positive.}
Hence, the field behaves as\begin{equation}
\phi (N)\approx [\phi _{e}^{2+p}-\alpha Np(2+p)]^{\frac{1}{2+p}}
\end{equation}
and the relevant expressions become\begin{equation}
\epsilon =\frac{\alpha ^{2}p^{2}}{2}[\phi _{e}^{2+p}-\alpha Np(2+p)]^{\frac{-2(1+p)}{(2+p)}}=2^{\frac{-(4+3p)}{2+p}}(\alpha p)^{\frac{2}{2+p}}\left( \frac{n-1}{1+p}\right) ^{\frac{2(1+p)}{2+p}}
\end{equation}
\begin{equation}
\label{eq:kinneyriottonmin1}
n-1=\frac{2\alpha p(1+p)}{\phi ^{2+p}}=\frac{2\alpha p(1+p)}{\phi _{e}^{2+p}-\alpha Np(2+p)}
\end{equation}
\begin{equation}
\label{eq:krdndlnk}
\frac{dn}{d\ln k}=\frac{-2\alpha ^{2}p^{2}(1+p)(2+p)}{[\phi _{e}^{2+p}-\alpha Np(2+p)]^{2}}=\frac{-(2+p)}{2(1+p)}(n-1)^{2}
\end{equation}
where we have the restriction \( \epsilon (N)<1 \). As long as \( \phi _{e} \)
is chosen such that\begin{equation}
\phi _{e}\gg [60\alpha p(2+p)]^{\frac{1}{2+p}}
\end{equation}
we can easily get greater than \( 60 \) efolds. As one can see, one
can obtain \begin{equation}
\frac{dn}{d\ln k}\sim -O((n-1)^{2})
\end{equation}
which can be large if \( (n-1)\approx 0.2 \) but the running does
not change sign to the red part of the spectrum.

From a field theoretic point of view, this potential is not very natural
because one generically expect terms of the form \( V_{0}\phi ^{2} \)
to spoil the potential. Not only must such term be less than \( V_{0} \),
but it must be less than \( V_{0}\alpha /\phi ^{p} \). Because of
this second constraint, one can easily show that even after introducing
a fine tuning in \( \alpha  \), \( \phi ^{2}>\alpha /\phi ^{p} \)
if one wants to have \( n-1\approx 0.2 \) at \( N=60 \).

\subsection{Wilson Line as an Inflaton}
The works of Refs.~\cite{extranatural1,extranatural2} propose 
an inflationary scenario in which
the inflaton is a Wilson line field \( \theta  \) of a 5 dimensional
\( U(1) \) gauge field whose fifth component \( A_{5} \) is integrated
around the extra 5th dimension. If \( \theta  \) is coupled to an
extra dimensional field charged under this gauge symmetry, the 4D
effective potential for \( \theta  \) reduces to
that of a pseudo-Nambu-Goldstone boson with an effective decay constant
of
\begin{equation}
f_{\rm eff}=\frac{1}{2\pi g_{4}R} \ ,
\end{equation}
 where \( g_{4} \) is the 4D effective gauge coupling constant. If
\( g_{4}\ll 1 \), then, even if \( R\gg 1 \) (in Planck units), \( f_{\rm eff}\gg 1 \)
can be arranged, thereby alleviating the usual problems associated
with natural inflation \cite{natural},
namely that of making one of the slow-roll
parameters
\begin{equation}
\eta \sim \frac{1}{f_{\rm eff}^{2}}
\end{equation}
much smaller than unity. The great advantage of this scenario is
that, due to the nonlocal nature of the field \( \theta  \) and gauge
invariance, the effective potential is protected from quantum
corrections that can spoil the inflationary slow-roll. 

Based on this scenario, the authors of Ref.~\cite{Feng:2003mk} proposed a method
of obtaining a large running. They introduce one massive and one massless
fermion in the bulk, charged under the extra-dimensional \( U(1) \)
gauge group, with charges \( q_{1} \) and \( q_{2} \), respectively,
giving rise to an effective potential of the form
\begin{eqnarray}
V(\theta ) & = & \frac{3}{64\pi ^{6}R^{4}}\sum _{n=1}^{\infty }\frac{1}{n^{3}}\left[ \frac{\cos (nq_{1}\theta )}{n^{2}}+\right. \nonumber \\
 &  & \left. e^{-n2\pi RM_{2}}\left( \frac{(2\pi RM_{2})^{2}}{3}+\frac{(2\pi RM_{2})}{n}+\frac{1}{n^{2}}\right) \cos (nq_{2}\theta )\right] \label{eq:extradimraw} \ ,
\end{eqnarray}
where \( M_{2} \) is the mass of the massive fermion. Keeping only
the \( n=1 \) term, redefining \( \theta =\phi /f_{\rm eff} \) and adding
a constant term, they obtain the potential
\begin{equation}
V=V_{0}\left[ 1+E(\sigma )-\cos \left(\frac{q_{1}\phi }{f_{\rm eff}}\right)-\sigma 
\cos\left(\frac{q_{2}\phi }{f_{\rm eff}}\right)\right] \ , 
\label{eq:zhangpot}
\end{equation}
where \( 1+E(\sigma )\sim O(1) \) is a term independent of the inflaton
field \( \phi  \) added to make the potential vanish at the minimum (The \( \sigma  \)
and \( V_{0} \) definitions trivially follow from matching). When
\( E(\sigma )=\sigma =0 \), the potential is the usual natural inflationary
potential. To obtain a large running, the authors have, by adding the massive
fermion, introduced \( \sigma \ll 1 \) dependent terms which modulate
the original potential with small amplitude wiggles (when \( q_{2}\gg q_{1} \)).
The running they attain is of order \( (n-1)^{3/2}\sqrt{\epsilon /\sigma } \)
which, because of phenomenological restrictions has an order of magnitude
limit of \( (n-1)^{2} \), similar to Ref.~\cite{Kinney:1998dv}. However,
the model defined in Eq.~(\ref{eq:zhangpot}) has the advantage of being able to run from blue to
red.

To see how the
running occurs, consider the slow-roll parameters. Expanding in powers of
\( \sigma  \) in the limit that \( \sigma q_{1,2}\ll 1 \), while
\( \sigma q_{2}^{2}\sim O(1) \) (and \( q_{1}\ll q_{2} \)), we
find 
\begin{equation}
\epsilon \sim \frac{1}{2\left(1-\cos \frac{q_{1}\phi }{f_{\rm eff}}\right)^{2}}\frac{q^{2}_{1}}{f_{\rm eff}^{2}}\sin ^{2}\frac{q_{1}\phi }{f_{\rm eff}}+O(\sigma q_{2}) \ ,
\end{equation}
\begin{equation}
\label{eq:etalimit}
\eta \sim \left[ \frac{\cos \frac{q_{2}\phi }{f_{\rm eff}}}{1-\cos \frac{q_{1}\phi }{f_{\rm eff}}}\right] \left(\frac{q_{2}}{f_{\rm eff}}\right)^{2}\sigma +\frac{q_{1}^{2}}{f_{\rm eff}^{2}}\left[ \frac{\cos \frac{q_{1}\phi }{f_{\rm eff}}}{1-\cos \frac{q_{1}\phi }{f_{\rm eff}}}\right] +{\cal O}\left((\sigma q_{2})^{2},\sigma \frac{q_{1}^{2}}{f_{\rm eff}^{2}}\right) \ ,
\end{equation}
\begin{equation}
\xi =\frac{-1}{\left(1-\cos \frac{q_{1}\phi }{f_{\rm eff}}\right)^{2}f_{\rm eff}^{4}}\left( q_{1}^{4}\sin ^{2}\frac{q_{1}\phi }{f_{\rm eff}}+\sigma q_{1}q_{2}^{3}\sin \frac{q_{1}\phi }{f_{\rm eff}}\sin \frac{q_{2}\phi }{f_{\rm eff}}\right) +O(\sigma q_{2}) \ .
\end{equation}
Away from any special points in \( \phi  \), we require
\begin{equation}
\epsilon \sim \frac{q_{1}^{2}}{f_{\rm eff}^{2}} \ ,
\end{equation}
\begin{equation}
\eta \sim \frac{q_{2}^{2}}{f_{\rm eff}^{2}}\sigma -\epsilon \ ,
\end{equation}
and
\begin{equation}
\xi \sim -\left(\frac{q_{1}^{4}}{f_{\rm eff}^{4}}+\sigma \frac{q_{1}q_{2}^{3}}{f_{\rm eff}^{4}}\right)\sim -\left( \epsilon ^{2}+(\eta +\epsilon )\frac{q_{1}q_{2}}{f_{\rm eff}^{2}}\right) 
\end{equation}
all to be small. Furthermore, \( \eta  \) must not change
sign more than once within about \( 10 \) efolds. This requires
\begin{equation}
\frac{q_{2}}{f_{\rm eff}}(\Delta \phi )\sim 10\sqrt{2\epsilon (\eta +\epsilon )/\sigma }\sim 10\sqrt{2}\frac{q_{1}q_{2}}{f^{2}_{\rm eff}}<\pi \ .
\end{equation}
Finally, to have 60 efolds, we must have
\begin{equation}
N\sim \frac{(\pi /2)f_{\rm eff}}{q_{1}\sqrt{2\epsilon }}\sim \frac{\pi }{2\sqrt{2}\epsilon }\sim 60 \ ,
\end{equation}
which requires
\begin{equation}
\epsilon <0.02 \ .
\end{equation}
Since \( \eta  \) must cancel against \( \epsilon  \) to run from
blue to red, it must be true that \( \eta <4\epsilon \sim 0.08 \).
Hence, without fine tuning initial conditions, one expects a maximum
running of order \( (n-1)^{3/2}\sqrt{\epsilon /\sigma }\sim |\eta q_{1}q_{2}/f_{\rm eff}^{2}|<0.02. \)
A more careful analysis by Ref.~\cite{Feng:2003mk} indicates that even with fine
tuning of the initial conditions, the running is of the order \( (n-1)^{2} \)
(note that if the initial conditions are fine tuned such that \( q_{1}\phi /f_{\rm eff} \)
is close to \( \pi  \), \( \epsilon  \) is suppressed and therefore
tensor perturbations are suppressed).

For completeness, we write the formula for \( n-1 \) corresponding
to this model in the limit prescribed for Eq. (\ref{eq:etalimit})
as\begin{equation}
\label{eq:nmin1renjie}
n-1\approx \frac{q_{1}^{2}}{f_{\rm eff}^{2}}\left(1-2\csc ^{2}\frac{q_{1}\phi }{2f_{\rm eff}}\right)+\sigma \frac{q_{2}^{2}}{f_{\rm eff}^{2}}\cos \frac{q_{2}\phi }{f_{\rm eff}}\csc ^{2}\frac{q_{1}\phi }{2f_{\rm eff}}.
\end{equation}

\end{document}